# DAS for 2D MASW Imaging: A Case Study on the Benefits of Flexible Sub-Array Processing

Michael B. S. Yust[1*], Brady R. Cox[2], Joseph P. Vantassel[3], and Peter G. Hubbard[4]

*(1) The University of Texas at Austin*
*(2) Utah State University*
*(3) Virginia Tech*
*(4) University of California, Berkeley*
*\*corresponding author: yustm@utexas.edu*

## Abstract

Distributed acoustic sensing (DAS) is a relatively new technology for recording stress wave propagation, with promising applications in both engineering and geophysics. DAS's ability to simultaneously collect high spatial resolution data (e.g., 1-m channel separation) over long arrays (e.g., kilometers) suggests that it is especially well suited for near-surface imaging applications such as 2D MASW (multi-channel analysis of surface waves). 2D MASW aims to produce a pseudo-2D cross-section of shear-wave velocity ($V_S$) for the purpose of identifying and characterizing subsurface layering and anomalies. These pseudo-2D $V_S$ cross-sections are produced by spatially interpolating numerous 1D $V_S$ profiles extracted from overlapping sub-arrays along the testing alignment. When using traditional seismic equipment (e.g., geophones and 24-channel seismographs), these sub-arrays are typically collected in a roll-along configuration, where the equipment is continuously moved along the alignment at some predetermined sub-array interval. In contrast, DAS does not suffer from the same limitations, as data from all shot locations are simultaneously recorded along the entire length of the DAS array at a constant channel separation. This alleviates the requirements to pre-determine sub-array length and sub-array interval during the data acquisition stage, and allows for multiple sub-array geometries to be investigated during the processing stage. The present study utilizes DAS data collected at high spatial resolution to evaluate the effects of sub-array length on 2D MASW results at a single, well-characterized field test site. We organize the DAS waveforms into multiple sets of overlapping MASW sub-arrays of differing lengths, ranging from 11 m to 47 m, allowing for direct comparison of the derived pseudo-2D $V_S$ results at the same site. We show that the length of the individual MASW sub-arrays has a significant effect on the resulting $V_S$ cross-sections. In particular, there is a tradeoff between long sub-arrays that improve dispersion data quality and allow deeper characterization and shorter sub-arrays that improve lateral resolution. We also show that sub-array length has a significant impact on the resolved location of large impedance contrasts at our study site and evaluate those locations compared to invasive testing. Our results suggest that *a priori* information should be used, when possible, to select the optimal sub-array length for 2D MASW analyses to address project-specific goals. If *a priori* information is not available, the analysts may need to consider multiple sub-array geometries, as is made possible by DAS, to properly evaluate the uncertainty of 2D MASW results. This study demonstrates that DAS is capable of collecting data for 2D MASW in a manner that is efficient, flexible, and pragmatic.

## Introduction

Surface wave methods are powerful tools for non-invasive seismic site characterization. One of the most popular testing methods is the multichannel analysis of surface waves (MASW), capable of producing a 1D shear-wave velocity ($V_S$) model of the subsurface (Park et al. 1999, Foti 2000). MASW is traditionally performed using a linear array of geophones and one or more 24-channel seismographs to record surface waves generated by an active source (i.e., hammer, weight drop, or vibroseis shaker) located off one or both ends of the array. As the number of geophones is often fixed by the availability of equipment, the analyst must balance the finer vertical layer resolution provided by smaller receiver spacings with the greater characterization depth provided by a longer array (Foti et al. 2018). Soon after the introduction of MASW, engineers began to explore how this new method could be used to characterize 2D variations of subsurface $V_S$ (Miller et al. 1999). Xia et al. (2000) proposed the use of MASW to construct pseudo-2D $V_S$ cross-sections by combining multiple 1D $V_S$ profiles resulting from multiple MASW datasets collected along a common alignment. This approach came to be known as 2D MASW.

2D MASW surveys typically use the roll-along method (Mayne 1962), with 24 or 48 geophones mounted on a land streamer system at a fixed receiver spacing. The number of geophones and the choice of receiver spacing pre-determines the length of the sub-array used during data acquisition and processing. The land streamer is pulled behind a vehicle, allowing the geophone array to be moved along the survey alignment and incrementally stopped at a predetermined horizontal distance called the sub-array interval. The sub-array interval is typically set equal to some portion of the total sub-array length (e.g., 1/4 or 1/3), such that there is significant spatial overlap between adjacent sub-arrays. The sub-array interval also determines the horizontal distance between the 1D $V_S$ profiles that will ultimately be interpolated to obtain the pseudo-2D $V_S$ image. Typically, a source such as a weight drop is attached to the towing vehicle to generate the active shots. In the interest of rapid data acquisition, generally only one shot location with a fixed offset is used for each MASW sub-survey Even when data is collected using larger, mobile or stationary geophone arrays, the traces are often reorganized to mimic the roll-along method with a single shot location (Park & Miller 2005a, 2005b, Thitimakorn et al. 2005).

The 2D MASW process has been used successfully on a variety of near-surface imaging tasks. In their initial paper, Xia et al. (2000) were able to successfully identify multiple subsurface features, including a bedrock channel in Olathe, Kansas, which was later confirmed with drilling, and a known steam tunnel at the University of Kansas. Thitimakorn et al. (2005) utilized 2D MASW to survey a 1950 m segment of Interstate 70 in St. Louis, Missouri. They used 12 4.5 Hz vertical geophones at a 3-m spacing with 12-channel sub-arrays (33 m) to construct a pseudo-2D $V_S$ cross-section with a sub-array interval of 12 m between 1D $V_S$ profiles. Based on this 1950-m-long cross-section, Thitimakorn et al. (2005) were able to identify depths to bedrock ranging from 6 m to 13.4 m, which agreed well with 19 boreholes drilled along the testing alignment. Park & Miller (2005a, 2005b) performed 2D MASW at 84 sites near Lawton, Oklahoma and 10 sites in Kansas to check for voids or other weak areas. They performed multiple 2D MASW surveys at each location with three parallel alignments and used a fourth alignment perpendicular to and bisecting the other three to ensure the site was characterized as thoroughly as possible. They used 48 stationary geophones at a 1.22-m spacing which were recompiled to simulate 24-channel roll-along acquisition with a 1.22-m sub-array interval. Mohamed et al. (2013) performed 24 collocated P-wave refraction and 2D MASW surveys at a site outside of Cairo, Egypt. They used 13 sub-arrays consisting of 24 geophones with 1-m spacings and a sub-array interval of 4 m to collect data at 24 sites. Mohamed et al. (2013) found that the 2D MASW surveys were more effective than P-wave refraction at detecting near-surface anomalies and low-velocity layers. The 2D MASW cross-sections identified low-velocity regions that, when boreholes were drilled at the site, were confirmed to correspond to claystone layers experiencing swelling due to nearby water sources, including irrigation and a swimming pool. Ismail et al. (2014) performed both 2D MASW and shear-wave reflection profiling along two alignments in southern Illinois totaling 3.7 km. They used sub-arrays consisting of 48 geophones with 1.5-m spacings and a sub-array interval of 7.5 m. They were able to map the depth to bedrock, including identification of near-surface faults, with the results from both methods agreeing well. Ismail et al. (2014) noted that, while the reflection survey was better able to resolve thin near-surface layering in the unconsolidated sediments above bedrock, the 2D MASW surveys provided better estimates of $V_S$ and were easier to perform. While not exhaustive by any means, the above-cited studies are indicative of successful applications of 2D MASW using traditional equipment and the standard roll-along method with a single, predetermined survey geometry (i.e., number and spacing of geophones, sub-array length, sub-array offset interval, shot location, etc.). However, despite its successful use in many projects, the potential of 2D MASW is still limited by practical constraints which require the analyst to determine that geometry prior to the data acquisition stage. This significantly reduces the options available during the processing stage and can have a significant impact on the survey results.

Multiple studies of synthetic data (Park 2005, Mi et al. 2017, Crocker et al. 2021, Arslan et al. 2021) have found that array geometry, especially array length, has a significant impact on the vertical and horizontal resolution of 2D MASW and its ability to accurately resolve layer boundaries and $V_S$ anomalies in the subsurface. Additionally, Yust (2018) demonstrated that using only a single shot location can result in misinterpretation of dispersion data if significant higher-mode energy is present. While using multiple shot locations helps minimize the risk of mode misidentification, it is not easy to implement using the roll-along method with traditional equipment. Thus, 2D MASW could be improved if a single, predetermined survey geometry did not have to be specified at the data acquisition stage, which would allow for greater flexibility at the data processing stage. This study aims to demonstrate how distributed acoustic sensing (DAS) can be used to collect field data for 2D MASW without the restrictions of traditional geophone arrays, allowing for greater flexibility in data acquisition and processing. Specifically, we examine how changing the length of each MASW sub-array, which is trivial for 2D MASW data collected using DAS, affects the resulting $V_S$ cross-sections at a well-characterized site. However, before the testing performed in this study can be fully discussed, it is important to first cover additional background information about how traditional 2D MASW testing is performed, such that modifications to the traditional approach discussed later in the paper can be better understood.

**Traditional 2D MASW**

Traditional MASW testing consists of three main steps: acquisition of surface-wave data in the field, processing the collected records to extract experimental dispersion data, and inverting the experimental dispersion data to produce a 1D subsurface $V_S$ model (Foti et al. 2015). The inverted 1D $V_S$ model is typically assumed to best represent the 1D layering and material properties beneath the center of the MASW array. However, due to the nature of the dispersion processing and inversion stages, which are inherently 1D, the $V_S$ profile represents a lateral average of all the materials beneath the array. 2D MASW follows these same three steps and adds a fourth one: combining multiple 1D $V_S$ profiles to create a pseudo-2D velocity cross-section along the testing alignment. Therefore, 2D MASW requires many adjacent MASW surveys to produce the 1D $V_S$ profiles that allow for pseudo-2D interpolation. Due to the large amount of data that needs to be collected (e.g., hundreds of shot records), the focus when performing data acquisition in the field for 2D MASW analysis is generally on efficiency. Hence, the roll-along method described above is typically utilized.

As the processing and inversion of 2D MASW data for each sub-array follows that of MASW, the following discussion will focus on only those aspects that are of particular importance. The interested reader is referred to the following references for more information (Park et al. 1998, Park et al. 2007, Foti et al. 2015, Foti et al. 2018, Vantassel & Cox 2022). Dispersion processing of surface wave data can be performed on the raw recorded wavefield or, if using a sweeping source, the wavefield cross-correlated with the source (Xia et al., 2000). The recordings can then be transformed to the frequency-wavenumber domain using various wavefield transformations (e.g., Nolet & Panza 1976, McMechan & Yedlin 1981, Park et al. 1998, Zywicki 1999, Xia et al. 2007, Luo et al. 2009a). For many datasets, the referenced transformations produce similar, although typically not identical, estimates of surface wave dispersion (Foti et al. 2015, Rahimi et al. 2021, Vantassel & Cox 2022). As such the MASW transform can be considered a source of dispersion uncertainty (Vantassel & Cox 2022). If multiple shot locations are used for each sub-array, differences in surface wave dispersion may also energy, these differences can also be quantified as part of the site-specific dispersion uncertainty (Cox & Wood 2011, Vantassel & Cox 2022). However, due to the additional time and effort required to use multiple transformations and multiple shot offsets, 2D MASW acquisitions typically use a single transformation and a single shot location and do not quantify dispersion uncertainty.

Once the experimental dispersion data has been obtained through wavefield processing a model of the subsurface is inferred through inversion (i.e., solving an inverse problem). Briefly, the inversion seeks to find the 1D ground model whose theoretical dispersion data best fits the experimental dispersion data measured in the field (Park et al. 1998). The 1D ground model in the inversion is defined by a set number of layers each defined by their thickness, mass density, compression-wave velocity, and $V_S$. The inverse problem is challenging as it is ill-posed, non-linear, and mixed-determined, with no guarantee of a unique solution (Foti et al. 2015). To find the ground models whose theoretical dispersion best fits the experimental data there are two main search approaches: gradient-based and gradient-free. Gradient-based methods, also referred to as local-search methods, rely on an initial starting model and the gradients of the misfit function (typically an L2 norm) with respect to each of model's parameter to converge to a local minimum through an iterative process (Socco et al. 2010). Gradient-free methods, also referred to as global-search methods, instead rely on sampling the model solution space to find the model(s) that best fit the data. These searches can be purely random (earlier samplings do not affect later samplings) or adaptive (later samplings learn from previous samplings). As local-search methods are faster than global-search methods (i.e., they typically require the solution of fewer forward problems) they have been used in the vast majority of previous 2D MASW studies (Foti et al. 2018). However, local-searches are known to be less rigorous then global-search and as a result are susceptible to becoming stuck in sub-optimal solutions (i.e., local minima and saddle points in the space of the inversion objective function) and should be used cautiously in environments where an accurate starting model cannot be selected *a priori* (Socco et al. 2010).

In addition to the optimization algorithm, the inversion is also strongly influenced by the inversion's parameterization (i.e., the number of assumed layers and the upper and lower limits of each parameter). Most 2D MASW studies consider only a single layering parameterization consisting of many layers with fixed thickness. These layers are often of uniform thickness, but may increase in thickness with depth (Xia et al. 2000). While the use of many layers may increase the ability of the inversion algorithm to fit the target dispersion data, it can also result in unrealistic $V_S$ profiles with large changes in velocity over short depth intervals, especially when velocity reversals are allowed at all layer boundaries (Song et al. 2020; Crocker et al. 2021). To address these issues, parameterization methods such as the layering ratio (Cox & Teague 2016) and layering by number (Vantassel & Cox 2021) approaches can be used to systematically investigate the sensitivity of the inversion to the choice of layering parameterization.

Once the inversion process has produced a 1D $V_S$ profile (or several 1D $V_S$ profiles if different inversion layering parameterizations are considered and uncertainty is acknowledged) for each MASW sub-array, those profiles can then be combined into a pseudo-2D $V_S$ cross-section. This is done by placing each

1D profile at the lateral position corresponding to the middle of the respective MASW sub-array. The $V_S$ values between those positions can then be interpolated (Xia et al. 2000). Luo et al. (2009b) examined the assumption that the 1D $V_S$ profiles were located at the midpoint of the MASW sub-array and found it to be reasonable. They also found that the dispersion data extracted from MASW testing is primarily affected by the subsurface conditions under the receiver spread itself and not under the space between the receivers and the shot location. As the primary goal of most 2D MASW surveys is to identify subsurface features such as low stiffness zones and layers with high impedance contrast, the lateral resolution of the pseudo-2D cross-section is a critical part of the analysis. Park (2005) examined the impact of sub-array length and sub-array interval using synthetically generated waveforms. Park found that the array length should be balanced between maximizing length, to improve dispersion data quality and maximize characterization depth, and minimizing length to reduce the amount of spatial averaging that occurs within each sub-array. This spatial averaging caused smearing of the subsurface details, resulting in reduced lateral resolution for longer arrays. Park (2005) also found that the sub-array interval should be kept below the sub-array length and that there should be some overlap between successive sub-arrays. Mi et al. (2017), Arslan et al. (2021), and Crocker et al. (2021) evaluated the ability of 2D MASW to detect anomalous structures within the subsurface. Mi et al. (2017) analyzed a combination of synthetic and field data sets and concluded that anomalous velocities could not be accurately resolved for features shorter than the sub-array length. Arslan et al. (2021) and Crocker et al. (2021) utilized over 3,000 synthetic data sets to examine the detection and resolution abilities of MASW depending on anomaly size and depth. They found that anomalies less than half the length of the sub-array were unlikely to be detected. They also cautioned against blind application of 2D MASW to detect anomalies due to the inherently 1D nature of all surface wave methods. Despite the important influence of sub-array length on characterization depth, dispersion data quality, and lateral resolution, its effects on 2D MASW results have not been extensively studied. Due to the impracticality of adjusting geophone sub-array lengths in the field using traditional MASW equipment, these studies have largely been limited to synthetic data sets. In response, this study aims to examine the effects (i.e., data quality, vertical and lateral resolution of layer boundaries, etc.) of sub-array length by comparing pseudo-2D $V_S$ cross-sections of a single, well-characterized field site using DAS with ground truth obtained from invasive methods.

**Distributed Acoustic Sensing for 1D and 2D MASW**

DAS is a relatively new technology for recording stress wave propagation, with promising applications in both engineering and geophysics (Daley et al. 2013, Lindsey et al. 2017, Spikes et al. 2019, Hubbard et al. 2021a). DAS uses light propagated through fiber-optic cables to collect data over very large scales (e.g., kilometers) while still maintaining very high spatial resolution (e.g., meters), a feat that is not possible with traditional sensing methods such as geophone arrays (Soga & Luo 2018). This is accomplished by measuring the change in length of sections of fiber-optic cable using backscatter interferometry (Hartog 2018). The DAS array itself has two major components; the interrogator unit (IU) which produces the source light and measures backscatter, and the fiber-optic cable which carries the light as a waveguide and acts as a distributed interferometer. As light pulses are sent down the cable by the IU, some of the light reflects back toward the IU in the form of Rayleigh backscatter (Nakazawa 1983). In quantitative DAS, backscattered light originating from two locations along a sensing cable are compared to determine the change in length of the cable between them. The distance between these reflection points is set by the IU and is known as the gauge length. The change in optical phase between two backscatter sources locations is used to calculate the change in length between those points and, by extension, the strain in the cable (Hubbard et al. 2022). The fiber-optic cable acts as a linear array of sensors where each gauge length is a sensor, referred to as a channel. The cable can be either laid across the ground surface (Spikes et al. 2019) or buried to improve coupling with the soil (Galan-Comas 2015, Vantassel et al. 2022). Importantly, unlike traditional geophones, where each one acts as a point measurement, each DAS channel is a distributed measurement over the gauge length. The first thing that must be considered when determining if DAS can successfully be used for 2D MASW is whether the data produced by DAS is a suitable alternative to that from traditional geophones arrays.

Many previous studies have demonstrated that DAS can successfully be used to collect high-quality data for various geophysical and engineering applications. Herein, we focus primarily on previous applications of DAS for active-source MASW to assess whether it could be successfully used for 2D MASW. Galan-Comas (2015), Cole et al. (2018), Song et al. (2020), Lancelle et al. (2021), and Vantassel et al. (2022) all utilized DAS for active-source 1D and 2D MASW surveys covering various lateral extents. Lancelle et al. (2021) collected MASW data from a DAS array in Garner Valley, CA, as well as geophone and accelerometer arrays collocated with various sections of the DAS array. Lancelle et al. (2021) found that the dispersion data extracted from all three receiver types using a modified MASW procedure agreed well overall, with excellent agreement at high frequencies. Lancelle et al. (2021) also inverted the DAS and accelerometer data and found that the resulting $V_S$ profiles agreed with both each other and a borehole log from the site. Galan-Comas (2015) performed a rigorous MASW survey using collocated 70-m long geophone and DAS arrays with multiple shot locations off each end of the arrays. The dispersion data extracted from both arrays agreed very well across all shared frequencies. The only significant difference between each data set was that the resolved DAS data had a maximum frequency of 23 Hz, while the geophone data extended up to 56 Hz.

Vantassel et al. (2022) successfully extracted multi-modal Rayleigh-wave dispersion data from DAS records of two co-located fiber-optic cables and a traditional geophone array. All three Rayleigh modes extracted from each cable had excellent agreement with the same modes extracted from geophone records. Vantassel et al. (2022) demonstrated that converting the DAS data to common engineering units with the geophones (i.e., proportional to particle velocity) was not necessary when using frequency-dependent normalization in the MASW processing, simplifying the analysis workflow for DAS data. The multi-modal dispersion data was inverted and the resulting $V_S$ profiles agreed well with each other, as well as with cone penetration testing (CPT) data and the anticipated subsurface geologic conditions at the site. Additionally, Vantassel et al. (2022) identified gauge length and channel spacing as the two critical parameters for consideration when using DAS to collect MASW data. The importance of channel spacing is not surprising, given it is an analog to geophone spacing, which has been shown to be the primary factor that controls the minimum resolvable wavelength, requiring at least two sensors per wavelength (Foti et al. 2018, Park et al. 1999). However, less-intuitively Vantassel et al. (2022) found that the DAS gauge length strongly limits the minimum resolvable wavelength. Due to the distributed nature of DAS channels, they are not able to accurately measure signals with wavelengths near and below the gauge length. As a result, the minimum resolvable wavelength of DAS is the longer of either two channel spacings or the gauge length. These findings explain the discrepancy between the DAS-derived and geophone-derived dispersion data collected by Galan-Comas (2015) and others.

Cole et al. (2018) and Song et al. (2020) both utilized DAS to perform 2D MASW surveys following the traditional roll-along approach. Cole et al. (2018) utilized an existing 5-km installed fiber along a rail line in the UK to collect data along an approximately 500-m long alignment with 50 shot locations spaced roughly 10 m apart using both DAS and a 24-channel geophone array on a land streamer. The geophone land streamer array had a 2-m spacing and was moved for every shot such that the maximum shot offset was 50 m, resulting in a roughly 40-m overlap between each geophone sub-array. The DAS array had a total length of 5 km with a channel spacing of 1 m. Each shot was recorded along the entire length of the DAS array and then an appropriate subset of channels was selected. Cole et al. (2018) found that they were able to record coherent wave energy out to a maximum offset of about 125 m, resulting in a roughly 115-m overlap between each DAS sub-array. They found that they were able to successfully extract dispersion data from both the DAS and geophones that could theoretically be inverted into a pseudo-2D cross-section, but did not proceed to the inversion stage. Song et al. (2020) collected 2D MASW data using DAS for a 500-m long cable with 26 active shot locations. For each shot location, Song et al. (2020) used the 30 DAS channels (2-m spacing) ranging from 40 to 100 m away to extract dispersion data. This data was then inverted to form a pseudo-2D $V_S$ cross-section of the study area. Song et al. (2020) also observed distortion of the extracted dispersion data for wavelengths shorter than about 11 m, which, given they used a gauge length of 10 m, is consistent with the findings of Vantassel et al. (2022) that DAS cannot resolve dispersion data at wavelengths near and below the gauge length. Notably, neither of these groups fully utilized the potential flexibility of DAS in their 2D MASW surveys, with both studies still following the traditional roll-along approach of selecting a single sub-array geometry relative to a single shot location.

One of the key advantages of performing 2D MASW with DAS is that no matter the length of the fiber optic cable, the sensor spacing and gauge length are constant, removing the tradeoff between sub-array length and sensor spacing imposed by traditional seismic equipment. Additionally, source shots from all shot locations are recorded simultaneously by every channel of the full DAS array. This means that the exact location and length of the individual MASW sub-arrays can be determined after field testing has been performed instead of before, creating the ability to change sub-array parameters like receiver/channel spacing and shot location during the processing stage. This study will take advantage of these strengths to demonstrate how DAS can be used to collect a wealth of 2D MASW data that allows for significant flexibility in processing much more quickly and efficiently than traditional geophone surveys.

Rigorous inversion techniques are used in this study to analyze DAS data collected along a 200-m long fiber optic cable with a channel spacing of 1.02 m and a gauge length of 2.04 m. This study also examines the effect of sub-array length on the resulting $V_S$ cross-sections using field data from the same site, something that is not practical using traditional geophone sub-arrays and has been limited to synthetic data or data collected at different sites in previous studies. In order to cover the full 200-m extent, we used 47 12-channel sub-arrays, 44 24-channel sub-arrays, and 38 48-channel sub-arrays, with six shot locations per sub-array (three off either end). Only 96 individual active-source shots (three shots at each of 32 locations) were needed to complete this survey as the same shot can be used for multiple sub-arrays. Attempting to complete the same survey using geophone sub-arrays would require 774 shots with 264 shots required to perform just the 24-channel tests. This efficiency, combined with the fact that DAS, as demonstrated by the studies discussed above, can collect high-quality surface wave dispersion data with the correct configuration (i.e. short channel spacing and gauge length), means that DAS is uniquely well suited for 2D MASW testing, outperforms geophone arrays, and will allow for future studies that would not be possible with traditional methods.

**DAS Testing Setup**

The DAS data for this study was collected at the NHERI@UTexas (Stokoe et al., 2020) Hornsby Bend test site in Austin, TX. An overview of the testing layout is shown in Figure 1. The endpoints of a 200-m long linear array were first established using a total station. Second, a trenching machine was used to excavate a 10-15 cm trench along the length of the array. Next, two fiber-optic cables, one manufactured by NanZee and one by AFL, were placed in the trench parallel to each other. A skid-steer loader was then used to backfill the trench and compacted the fill material to ensure good coupling between the cables and the surrounding soil. Finally, the two cables were spliced together at the far end of the array so that they could be recorded simultaneously, and the near end of the AFL cable was terminated to reduce end reflections. While splicing the cables together does allow for simultaneous interrogation of both cables, for simplicity, only the data collected using the NanZee cable was used in this study. Vantassel et al. (2022) and Hubbard et al. (2022), who used the same experimental installation, found no significant differences in the DAS data collected with the two different cables. The cable from NanZee Sensing Technologies (NZS-DSS-C02) is a single-mode, tightly buffered cable. This cable is specifically designed for strain sensing applications and has been shown to work well in a variety of civil engineering projects (Zhang et al. 2019, Hubbard et al. 2021a, 2021b). The tightly buffered optical fiber is surrounded by braided steel reinforcement to protect the fiber and increase the strength of the cable. A textured polyethylene sleeve encases the reinforced fiber and aids in the transfer of soil strains to the fiber.

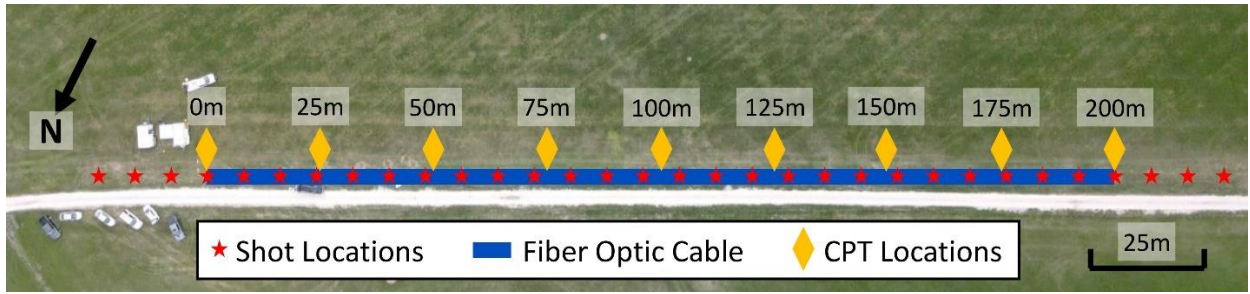

Figure 1: Aerial view of the Hornsby Bend test site showing the locations of CPT tests as well as the DAS fiber optic cable and the vibroseis shot locations.

The data analyzed in this study was collected approximately five months after the cable was initially installed. A temporary surface cable was used to connect the near end of the NanZee cable to the DAS IU in a nearby instrumentation trailer. The IU used in this testing was an OptaSense ODH4+ owned by NHERI@UTexas. Data was recorded continuously at a ping rate of 20 kHz and then decimated down to 1 kHz prior to dispersion processing. The ODH4+ is a variable gauge length IU, allowing the user to adjust the gauge length among multiple options for different applications. In this testing, the shortest possible gauge length of 2.04 m was selected. Additionally, the IU was configured with its minimum channel separation of 1.02 m. This results in a 50% overlap in the length of fiber being characterized for each channel. Unlike geophones, where the location of each channel is known, the exact location of each DAS channel is somewhat uncertain. While the spacing between channels is consistent, the absolute location of each channel is dependent on the exact length of every cable connected in series to the IU. A tap test was performed to determine which channels along the NanZee cable were associated with specific locations on the ground surface. The tap test consisted of lightly hitting the ground at known locations along the cable's length to see which channels had the maximum amplitude response relative to a given tapping location. In this study, the tap test results from 50, 100, and 150 m were used to determine the nearest DAS channel to each location, and the locations of all other channels were fixed relative to these measurements. This allowed the location of each channel to be determined with a maximum error of half the gauge length, which is sufficient for MASW testing. The tap test results from each end of the array were used to determine the range of usable DAS channels. A total of 196 usable channels were identified for the NanZee cable, resulting in a total array measurement length of 200.94 m. This DAS array is shown as a blue line in Figure 1.

The source used in this study was the Thumper vibroseis mobile shaker truck from the NHERI@UTexas experimental facility (Stokoe et al., 2020). Thumper is a moderate force shaker designed for testing in urban areas and has a maximum force output of about 27 kN. Thumper was used to produce a chirp signal with the frequency of shaking sweeping linearly from 5 to 200 Hz over 12 s with a 0.5 s cosine taper on each end. While Thumper can be configured to shake in either the vertical horizontal direction, only vertical shaking was used. Shaking was performed at 32 shot locations along the alignment of the array, beginning at -24 m and continuing every 8 m to 224 m. These shot locations are shown as red stars in Figure 1. Three sweeps were performed with Thumper at each shot location, resulting in a total of 96 shot records. A time break and the measured dynamic ground force output by Thumper were recorded for each sweep.

**Cone Penetration Testing**

Nine CPT soundings were performed adjacent to the fiber-optic cable at 25-m intervals from 0 to 200 m, as shown in Figure 1. The method developed by Robertson (2009) was used to determine the normalized soil behavior type index value ($I_c$) for all the CPT data collected. These results were then used to develop a soil-type cross-section of the subsurface conditions along the cable alignment, which is shown in Figure 2. The soil behavior type at each CPT sounding is shown as a color-coded column at the location of each sounding. Cohesive soil types have been colored blue and green, while granular materials are shown with tan, brown, and grey. Additionally, the measured tip resistance ($q_c$) profile from each sounding is shown with a red line and shaded area to the right of each column at a scale of 20 MPa between each column. When identifying sections of different soil behavior types for each column, vertical portions less than 10 cm long (approximately 6 samples) were ignored for simplicity. Based on the $q_c$ values and the soil behavior types, an interpretation of the significant layer boundaries beneath the site was made, as shown in Figure 2.

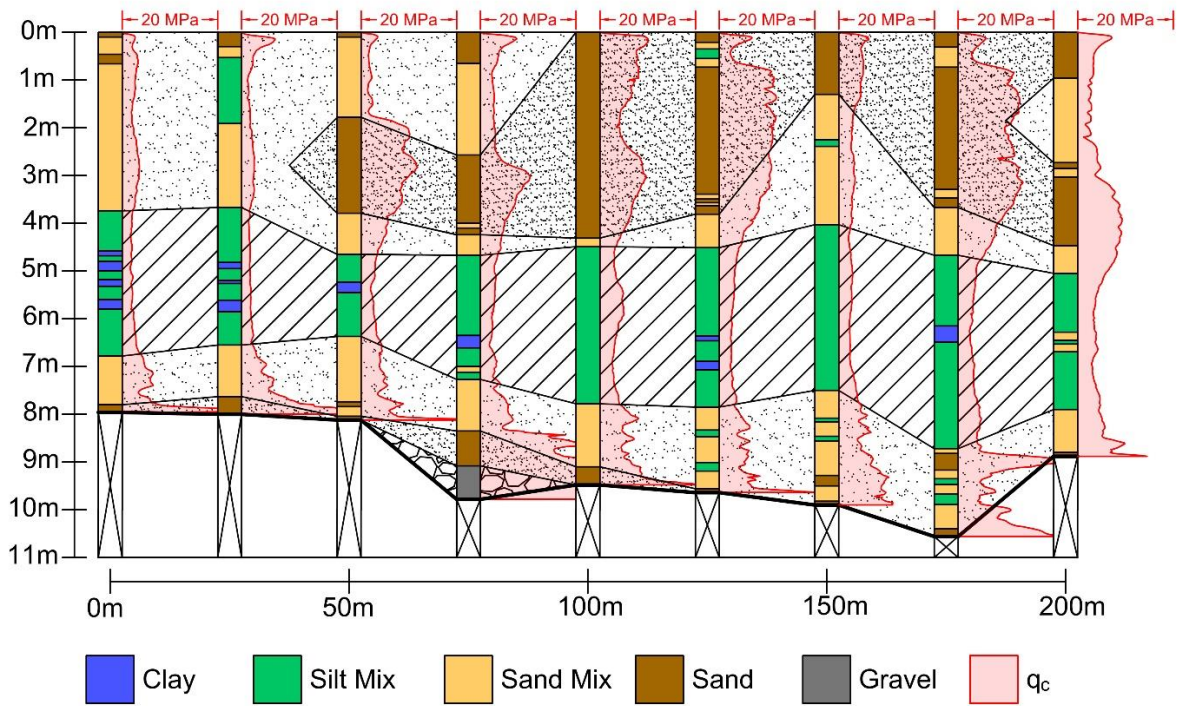

Figure 2: Cross-section of subsurface soil conditions beneath the DAS array based on CPT testing. At each CPT sounding location, the soil behavior type is indicated and the corresponding tip resistance ($q_c$) profiles are shown to the right of each sounding. An interpretation of the soil layering has been identified based on soil type and tip resistance.

The site appears to have 3 distinct layers above the depth of CPT refusal. The three major layers consist of a shallow (roughly 0 to 4 m) granular layer of sand and sand mix, an intermediate (roughly 4 to

7 m) cohesive layer of clay and silt, and a deeper (roughly 7 to 10m) granular layer of sand and sand mix. The granular layers are shown with a speckle pattern, while the cohesive layer is shown with diagonal hatching. Within the granular materials of the shallow layer, there appear to be some areas of greater strength, mostly near the middle and far end of the array. These regions are shown with a denser speckle pattern. Additionally, at the 75-m sounding location, it appears that the CPT was able to penetrate deeper into the stiffer material at the bottom of the deeper granular layer, which caused refusal in most other CPT soundings along the cable. The depth of refusal along the cable ranged from 7.96 to 10.56 m, with an average depth of 9.15 m. The depth of refusal generally trended deeper as the distance along the array increased, with the exception of the sounding at 75 m and the final sounding at 200 m.

In addition to standard soil behavior type classification, the CPT results were used to develop a preliminary pseudo-2D cross-section of the subsurface $V_S$ along the cable. To accomplish this, three CPT-to-$V_S$ correlations were used for each CPT sounding. The correlations used were those recommended by Wair et al. (2012) for their "All Soils" method. The all soils method was selected over the type-specific (sand or clay) methods, as using a combination of the type-specific methods produced unrealistic discontinuities in the resulting $V_S$ profiles. The three CPT-to-$V_S$ correlations recommended by Wair et al. (2012) were developed by Mayne (2006), Andrus et al. (2007), and Robertson (2009), and are shown as Equations 1, 2, and 3, respectively.

$$V_s = 118.8 \log(f_s) + 18.5 \tag{1}$$

$$V_s = 2.62 \, q_t^{0.395} I_c^{0.912} D^{0.124} SF \tag{2}$$

$$V_s = \left[10^{(0.55 I_c + 1.68)} (q_t - \sigma_v)/p_a\right]^{0.5} \tag{3}$$

These relationships determine $V_S$ as a function of a variety of factors including sleeve friction ($f_s$), corrected tip resistance ($q_t$), soil behavior type index ($I_c$), depth (D), total vertical stress ($\sigma_v$), and atmospheric pressure ($p_a$). Wair et al. (2012) modified each of the equations to use consistent units, with $f_s$, $q_t$, and $\sigma_v$ in kilopascals, D in meters, and $p_a$ = 100 kPa. The Mayne and Robertson relationships were developed for Quaternary soils in general, while the Andrus et al. relationship uses an age scaling factor (SF) to distinguish between Holocene (SF = 0.92) and Pleistocene (SF = 1.12) soils. Because the exact age of the soils at the Hornsby Bend test site is not known, an average value of SF = 1 was used to represent all Quaternary soils, as recommended by Wair et al. (2012). The resulting $V_S$ values from each of these relationships were averaged together to produce a single, mean $V_S$ profile for each CPT sounding. These mean $V_S$ profiles were then discretized at a 0.1-m intervals using linear interpolation first in terms of depth and then laterally between CPT soundings. The resulting pseudo-2D $V_S$ cross-section is shown in Figure 3. Based on these results, the $V_S$ above refusal has an approximate range of 150 to 300 m/s. There does not appear to be a consistent, distinct difference in $V_S$ between the granular and cohesive layers identified in the CPT data. Instead, the velocities tend to be lower on the near end of the array with stiffer material in the middle and far end of the array. While these CPT-to-$V_S$ correlation results should not be interpreted as the "true" values of $V_S$ within the subsurface, they do provide a helpful point of comparison. As such, these correlated CPT-to-$V_S$ values ($V_{S,CPT}$) will be used as a point of reference when evaluating the 2D MASW results above the depth of refusal.

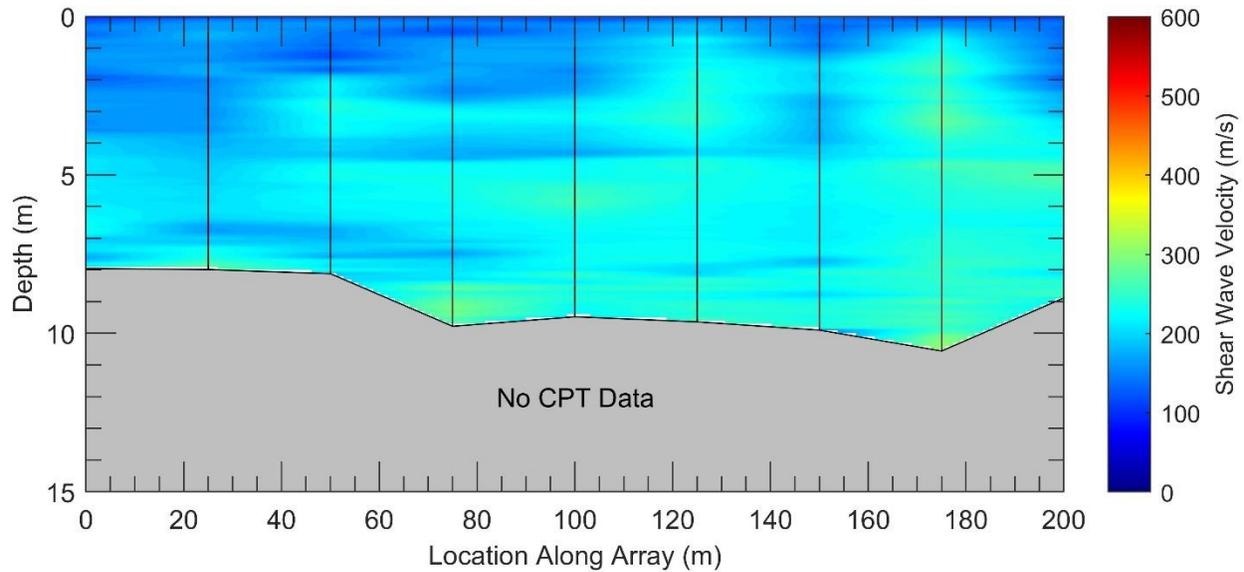

Figure 3: Pseudo-2D shear-wave velocity ($V_S$) cross-section developed using CPT-to-$V_S$ correlations and 9 CPT soundings spaced at 25-m intervals. The locations of the individual CPT soundings are shown with black vertical lines. Areas without any CPT data (i.e., below the depth of CPT refusal) are colored grey.

## 2D MASW Processing

Three different sets of MASW sub-arrays with increasing lengths were used in this study to examine the effects of sub-array length on the resulting 2D $V_S$ cross-sections. The first sub-array set consisted of 47 sub-arrays, with each sub-array using 12 channels. The second sub-array set consisted of 44 sub-arrays, each using 24 channels. The third sub-array set consisted of 38 sub-arrays, each using 48 channels. Although different numbers of channels were used for each sub-array set, all sets used a constant channel spacing of 1.02 m. Within each set, each sub-array was shifted by 4 channels (4.08 m) relative to the previous sub-array in the set for pseudo-2D MASW processing. Overall, a total of 129 separate MASW sub-arrays were analyzed in this study. As the sub-array interval between each sub-array is constant despite the number of channels in each sub-array increasing, the overlap between subsequent sub-arrays in a given set is significantly higher for the longer arrays. The 12-channel sub-arrays have a 67% overlap, which increases to 83% and 92% for the 24-channel and 48-channel sub-arrays, respectively. It is important to consider the overlap between subsequent sub-arrays in each set, as this can significantly affect the lateral resolution of the final 2D MASW results. Individual sub-array data sets were created from the DAS data by extracting sets of contiguous channels of the desired length while shifting 4 channels at a time down the array of 196 channels.

Once individual sub-arrays were extracted, the waveforms from each were processed separately to produce a single 1D $V_S$ profile located at the midpoint of that sub-array. The flexibility and efficiency of the DAS acquisition method allowed for more extensive and rigorous processing and inversion procedures than would have been practicable for most pseudo-2D MASW surveys. First, six different shot locations were considered for each individual sub-array. This was done by finding the three closest shot locations off either end of the sub-array that were greater than 4 m. These locations were selected so that a variety of shot offsets would be used, while also ensuring that the shot location was not too close to the first channel in the sub-array to mitigate near-field effects. Shot offsets for any given sub-array ranged from approximately 4 m at a minimum to approximately 24 m at a maximum. An example of one of the 24-channel sub-arrays is shown in Figure 4a, along with the locations of the six shots used in analyzing that sub-array. This sub-array will henceforth be referred to as the example sub-array. The example sub-array was the second 24-channel MASW sub-array along the DAS line and ranged from 4.08 m to 27.54 m. Once the shot locations were selected for each sub-array, the raw records from each shot location were correlated to the ground force output recorded by the vibroseis shaker truck for that shot, as recommended by Xia et al. (2000). The three correlated records from each shot location were then stacked in the time domain to improve the signal to noise ratio of the records. The correlated and stacked records for the -8 m shot location of the example sub-array are shown in Figure 4b.

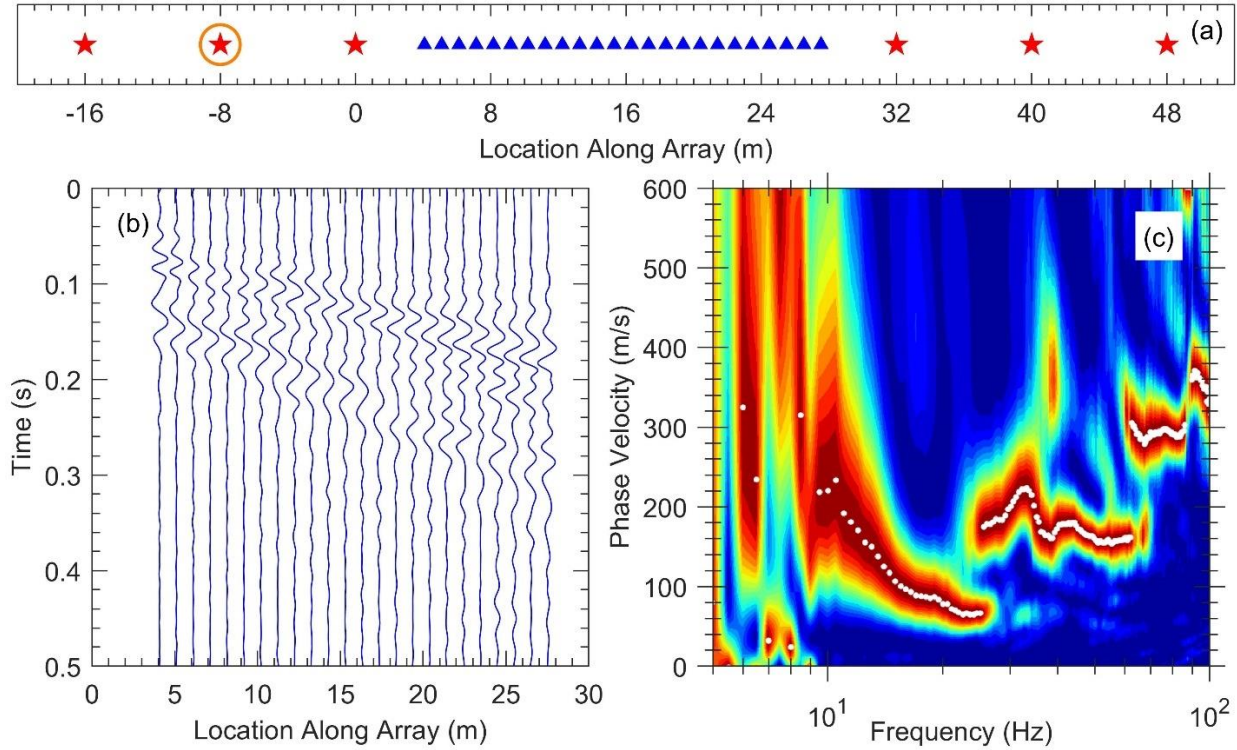

Figure 4: Example 24-channel MASW sub-array and the data extracted from it. (a) The location of DAS channels (blue triangles) and the 6 shot locations (red stars) analyzed for the sub-array. (b) Waterfall plot of the correlated and stacked vibroseis truck waveforms recorded with the sub-array for the highlighted shot location at -8 m. (c) Rayleigh dispersion image extracted from the waveforms for the highlighted shot location. The peak energy for each frequency value is marked with a white circle.

The MASW processing for each sub-array followed the workflow documented by Vantassel and Cox (2022) and utilized the *swprocess* Python package developed by Vantassel (2021). Each set of records was processed using the frequency-domain beamformer method with a cylindrical-wave steering vector and square-root-distance weighting (Zywicki & Rix 2005) to develop a Rayleigh dispersion image. The resulting dispersion image for the example records is shown in Figure 4c. This image shows the relative energy for different phase velocity-frequency pairs. The velocity value with the highest energy at each frequency was then selected automatically. Once all six shot locations for a given sub-array were processed, the dispersion data from each of them was combined to form a single dataset that could be used to calculate dispersion statistics. Prior to calculating dispersion statistics, the combined set of dispersion points was trimmed to remove obvious outliers and points that were determined not to correspond to the fundamental Rayleigh mode (R0). This combined set of dispersion points for the example sub-array is shown in Figure 5a, with different colors indicating the data from different shot locations. The data was then resampled in terms of frequency and the mean and standard deviation phase velocity of the points within each bin was calculated. The resulting R0 experimental dispersion data for the example sub-array is shown with black error bars in Figure 5a.

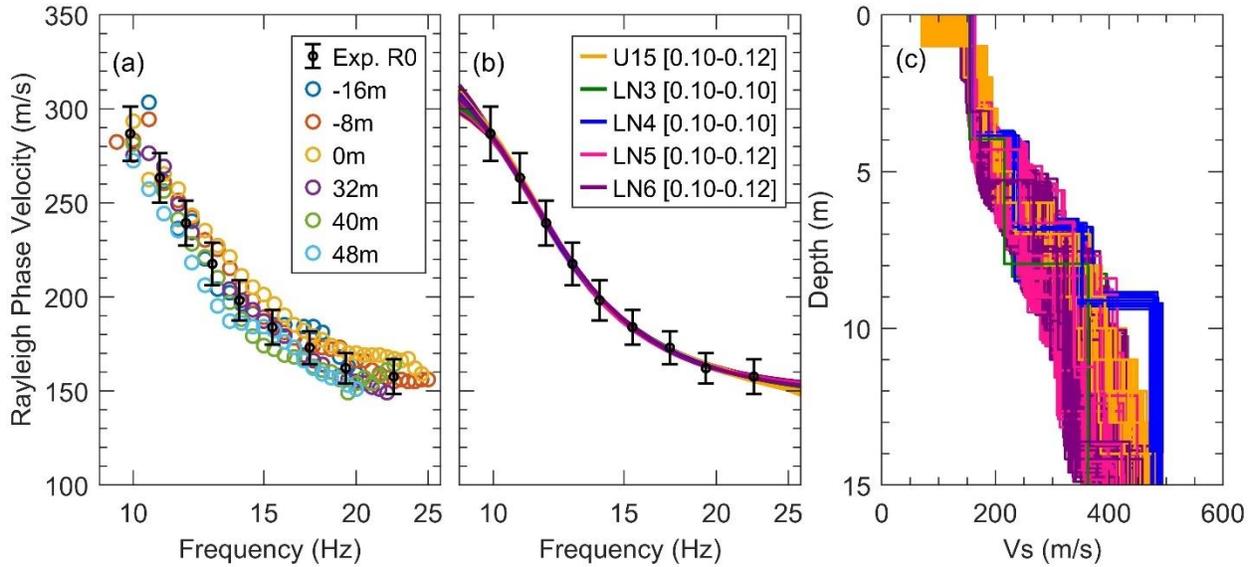

Figure 5: Example of the dispersion processing and inversion results for the example 24-channel sub-array. (a) Rayleigh dispersion data from all six shot locations after trimming and calculating statistics. (b) The fundamental Rayleigh mode (R0) experimental dispersion data and inversion-derived theoretical dispersion curves for the 300 lowest misfit trial models from each layering by number (LN) parameterization (1500 models total). The ranges of misfit values for the 300 best models from each parameterization are shown in brackets in the legend. (c) The 300 lowest misfit inversion-derived $V_S$ profiles from each parameterization.

The experimental dispersion data provides an opportunity to examine the effects of sub-array length on the MASW processing results. To illustrate this, all 129 sets of experimental dispersion data for the various sub-arrays are shown in Figure 6 in terms of both frequency and wavelength. Unsurprisingly, the length of each sub-array appears to have a significant impact on the wavelength range of the resolved dispersion data. While there is some variation between sub-arrays of the same length, the mean minimum and maximum wavelength values from each set of sub-arrays can be used to examine overall trends between them. The average minimum wavelength ($\lambda_{min,avg}$) resolved by each sub-array set was very consistent, with values of 6.6, 6.5, and 6.4 m for the 12-, 24-, and 48-channel sub-arrays respectively. The variation of the minimum wavelength between sub-arrays of the same length is also consistent across sub-array sets, with standard deviations ranging from 1.04 to 1.24 m. As the minimum resolvable wavelength is generally controlled by the sensor spacing within geophone arrays (Foti et al. 2018, Park et al. 1999), and given that all three sets of sub-arrays use the same channel spacing of 1.02 m and gauge length of 2.04 m, this agreement is unsurprising. The ability to maintain the same channel spacing and gauge length across sub-arrays of any length is one of the key strengths of utilizing DAS for 2D MASW, as it allows for data to be collected over long arrays without sacrificing the ability to resolve short wavelengths by increasing receiver spacing.

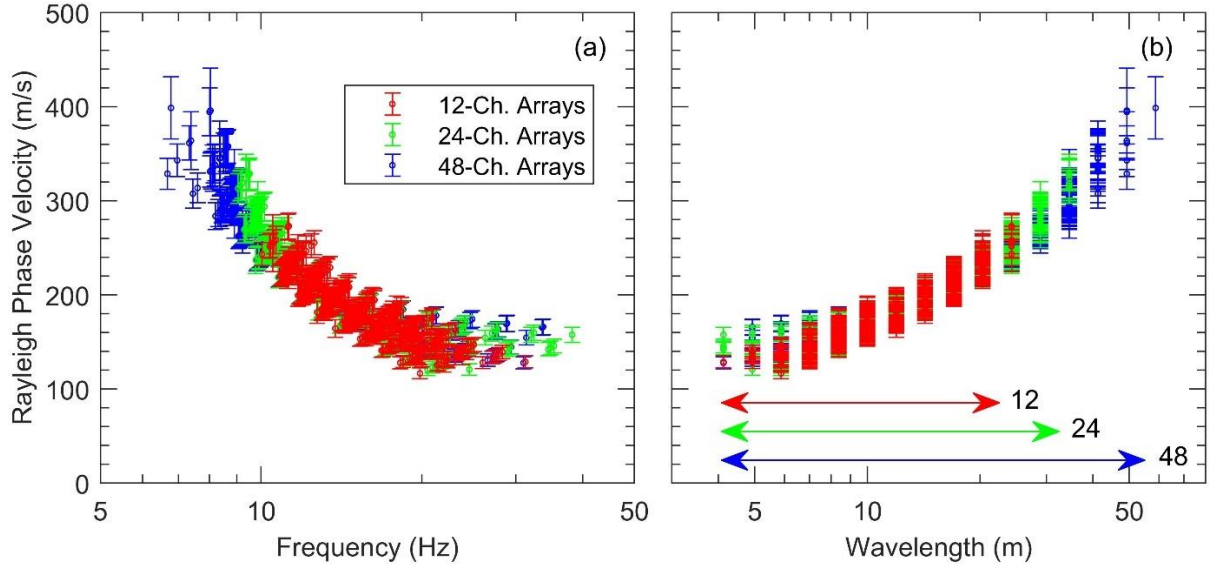

Figure 6: The fundamental-mode Rayleigh (R0) experimental dispersion data extracted for all 12-, 24-, and 48-channel MASW sub-arrays shown in terms of (a) frequency and (b) wavelength.

The key parameter where the dispersion data from sub-arrays of different lengths vary is the maximum resolved wavelength. The maximum resolvable wavelength ($\lambda_{a,max}$) for MASW testing is a function of both the shot location and the array length (L) (Vantassel & Cox 2021, Yoon & Rix 2009). If the shot location is too close to an array of a given length, near-field effects can cause phase velocities at longer wavelengths to be incorrectly measured (typically too low). Here, multiple shot locations were used for each individual sub-array to experimentally identify and mitigate these effects on the dispersion data. Consistent offsets were used for all three sets of sub-arrays to minimize variation between them. The exact effect of array length on maximum resolvable wavelength is disputed, with Foti et al. (2018) arguing that $\lambda_{a,max} \leq L$, while Park (2005) suggests a less restrictive value of $\lambda_{a,max} \leq 3L$. The maximum resolved wavelength for these datasets increased with array length, with average values ($\lambda_{max,avg}$) of 20, 28, and 39 m for the 12-, 24-, and 48-channel sub-arrays respectively. These values range from 0.6L to 2.2L. The maximum resolved wavelength is especially important, as it controls the depth to which the subsurface can be accurately characterized. For example, it is common to assume the maximum depth of the inverted $V_S$ profile is equal to one-half to one-third of $\lambda_{a,max}$ (Foti et al. 2018). Concerned only with characterizing as deep as possible, some analysts might be inclined to use as long an array as possible to try to maximize this value, but it is important to also consider the near-surface resolution of the results. Park (2005) shows that increasing the array length too much can cause significant distortion of subsurface features due to spatial averaging beneath the lateral extent of the array, while Mi et al. (2017) show that lateral resolution for any given array geometry decreases with depth, so identifying subsurface features at depth is doubly challenging. Therefore, it is important to balance the desired depth of characterization with the lateral resolution required to resolve structures of interest when selecting the length of each individual sub-array. Based on the results presented in Figure 6, it is presumed that the 48-channel sub-arrays will be better able to resolve deeper structure, however, they will also result in more spatial averaging of the subsurface properties. While the Hornsby Bend site considered in this study does not appear to have a large amount of lateral variability, the effects of spatial averaging can nonetheless still be observed in the dispersion data by examining the variability of the mean velocity values from different sub-array lengths. We examine this variability quantitatively by computing the standard error of the mean Rayleigh phase velocities for the 12-, 24-, and 48-channel sub-array lengths as 8.2, 7.1, and 6.1 m/s, respectively. As anticipated from the previous discussion, the 12-channel sub-arrays produced the most variable dispersion data (i.e., highest standard error in the mean) while the 48-channel sub-arrays produced the least variable data (i.e., lowest standard error in the mean). The decrease in standard error for the longer sub-arrays is indicating more spatial averaging for these sub-arrays than their shorter counterparts, and therefore that these arrays are less capable of observing smaller-scale variability across the Hornsby Bend site.

The experimental dispersion data for each sub-array was inverted to produce 1D shear-wave velocity profiles representing the subsurface conditions beneath the center of the sub-array. Five different layering parameterizations were considered during inversion as a means to account for epistemic uncertainty in the true subsurface layering (Cox & Teague 2016, Di Giulio et al. 2012). The five layering parameterizations used during inversion were the same for all 129 sub-arrays analyzed in this study. The first parameterization consisted of 14 layers with a uniform thickness of 1 m each overlying a half-space and is referred to as U15 (i.e., 15 total layers including the half-space). This parameterization was chosen because it is similar to those often used in 2D MASW analyses, where a large number of relatively thin, fixed-boundary layers are used during inversion (e.g., Seshunarayana & Sundararajan 2004, Park & Miller 2005a, Song et al. 2020). The other four parameterizations followed the layering by number (LN) method developed by Vantassel & Cox (2021) as part of the SWinvert workflow. This LN parameterization method allows for a predetermined number of layers within the $V_S$ model with the locations of layer boundaries free to find the optimal depths required to fit the dispersion data and limited only by the maximum inversion depth ($d_{max}$) and minimum layer thickness ($h_{min}$). LN parameterizations with 3, 4, 5, and 6 layers were used with $d_{max} = 15$ m and $h_{min} = 2$ m. The same maximum depth was used for all inversions regardless of the length of the sub-array and their corresponding $\lambda_{a,max}$ values to allow for comparison of all results to a consistent maximum depth (i.e., 15 m), despite the variation in the wavelength bandwidth of individual sets of dispersion data (refer to Figure 6). While this consistent maximum depth makes it easier to compare results from sub-arrays of different lengths, it is important to note that, based on the $\lambda_{max,avg}$ values reported above for the various sub-array lengths, resolved $V_S$ values will be less reliable at depths greater than about 10 m for the 12-m long sub-arrays.

The inversions in this study were performed using the Dinver module of the open-source Geopsy software (Wathelet et al. 2020), which uses a global neighborhood search algorithm. The inversions followed the SWinvert workflow and used the Dinver tuning parameters recommended by Vantassel & Cox (2021). Each inversion considered a total of 60,000 trial models. As recommended by Vantassel & Cox (2021), each inversion was performed three times to account for the variations that may result from the random starting seed of the search algorithm. Thus, a total of 15 inversions were performed for each sub-array (five parameterizations with three random starting seeds), resulting in 900,000 total trial models searched for each individual sub-array. The 1000 "best" (i.e., lowest misfit) trial models from each inversion were saved, from which the 300 with the lowest overall misfit were then selected for each parametrization. The 300 $V_S$ profiles for each layering parameterization used for inversion of the example array are shown in Figure 5c, with the corresponding theoretical dispersion curves shown in Figure 5b relative to the experimental dispersion data and the range of misfit values for each parameterization included inside brackets in the legend. These 300 $V_S$ profiles for each parameterization were then used to calculate a median $V_S$ profile for each layering parameterization.

**Results and Discussion**

The median $V_S$ profiles calculated from each inversion layering parameterization for each sub-array were used to produce pseudo-2D $V_S$ cross-sections of the subsurface beneath the 200-m long DAS array. Each median $V_S$ profile was first assigned a location along the DAS array equal to the midpoint of the corresponding MASW sub-array. The $V_S$ profiles were then discretized at a depth interval of 0.1 m down to the maximum inversion depth of 15 m to ensure a consistent number of data points (151) across all parameterizations. At each depth interval, the profiles were then linearly interpolated at an interval of 0.1 m from 0 m to 200 m, producing 2001 data points at each depth. Finally, a Gaussian filter ($\sigma = 10$) was applied to the entire grid of points to smooth out some of the abrupt and unrealistic discontinuities between profiles while still maintaining as much of the behavior identified in each individual profile as possible. The result is that each pseudo-2D $V_S$ cross-section consists of a 2001 by 151 grid of possible data points at common depth and lateral location values. As an example, the resulting $V_S$ cross-section for the U15 parameterization of the 24-channel MASW sub-arrays is shown in Figure 7. This pseudo-2D $V_S$ cross-section was generated from 44 individual 1D $V_S$ profiles with a total lateral extent of 175.44 m, ranging from 11.73 to 187.17 m (i.e., the mid-points of the first and last sub-arrays). All points in the grid outside of the first and last sub-array mid-points were set to a null value. Having this consistent grid of samples for all 2D cross-sections allows for comparisons between different inversion parameterizations, sub-array lengths, and the CPT-to-$V_S$ correlations.

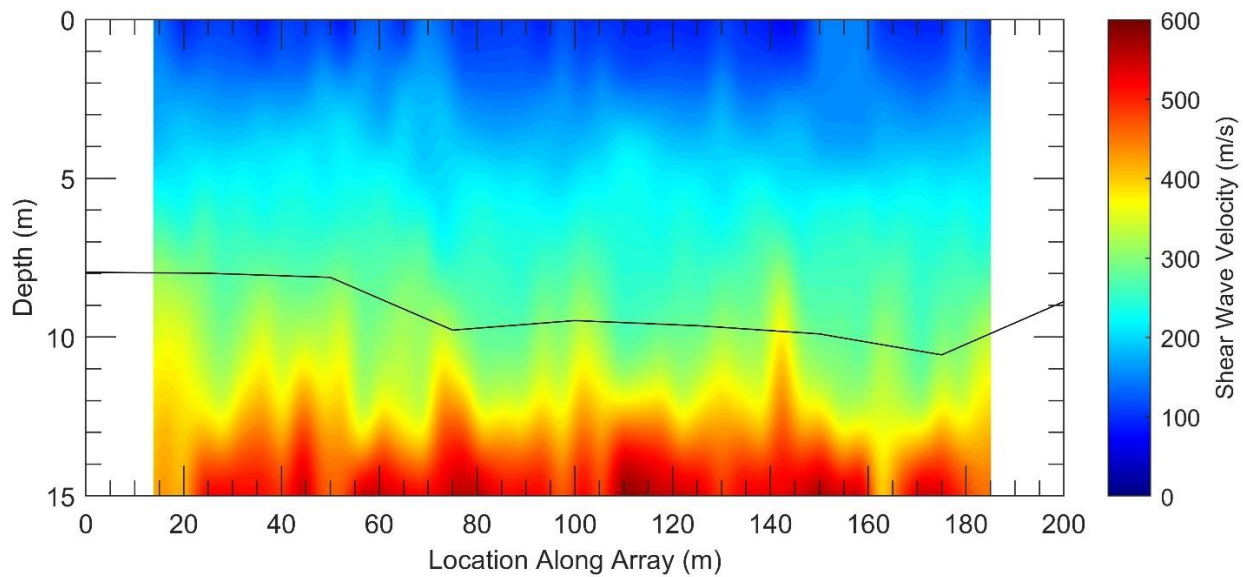

Figure 7: Pseudo-2D $V_S$ cross-section from the 44, 24-channel MASW sub-arrays inverted using the U15 parameterization. The depths of refusal for 9 CPT soundings along the array are shown with a solid black line.

The U15 results from all three sets of sub-arrays are shown in Figure 8. The pseudo-2D $V_S$ cross-sections for the 12- and 48-channels sub-arrays are shown in Figures 8a and 8e, respectively, with the 24-channel results from Figure 7 reproduced in 8c. The difference in lateral extent between the cross-sections is easy to see, with a maximum extent of 187.68 m for the 12-channel sub-arrays and 150.96 m for the 48-channel sub-arrays. While many 2D MASW surveys cover long distances, where this reduction in extent may be insignificant, for projects with tight lateral constraints it is important to consider this limitation when selecting the length of the sub-arrays. As shown in Figure 8, while all three cross-sections have similar half-space velocities, generally ranging from 450 to 600 m/s, the average depth at which higher velocities are reached varies significantly between each cross-section. For example, the depths where $V_S > \sim 400$ m/s from the 12-channel sub-arrays seem to correspond best with the depths of CPT refusal, and these depths consistently increase with each subsequent increase in sub-array length/number of channels. Specifically, the mean depths for $V_S$ exceeding 400 m/s are 11.89 m, 12.76 m, and 13.81 m for the 12-, 24-, and 48-channel cross-sections, respectively. However, results are more similar within the top third of all three cross-sections, with $V_S$ values transitioning from approximately 100 m/s at the ground surface to approximately 250 m/s at a depth of approximately 5 m. In the 12-channel cross-section, $V_S$ varies more significantly when moving laterally along the array, with transitions happening gradually at some locations and much more suddenly at others. In contrast, the $V_S$ transitions seen in the 48-channel cross-section are much smoother when moving laterally, with only a few isolated profiles varying significantly from those around them. As one might expect, the 24-channel cross-section shows behavior between the other two, with less variation than the 12-channel one and more than the 48-channel one. This lateral smoothing behavior for longer sub-arrays is consistent with the findings described by both Park (2005) and Mi et al. (2017). As the length of each sub-array is increased, more soil is being averaged within the MASW processing scheme. This results in a smearing effect, where $V_S$ variations of relatively small lateral extent can no longer be resolved and are simply merged with the surrounding materials. These results clearly indicate that, despite the phase velocities of the dispersion data shown in Figure 6 being consistent across sub-array lengths (at least in the zones of overlapping frequency/wavelength), changing the sub-array length can cause significant changes in the resulting $V_S$ cross-sections. Thus, some calibration to invasive field testing results may be needed in order to determine the best MASW processing and inversion scheme for a given application. For example, if one were looking to determine the depth to CPT refusal across this site, using a criteria of $V_S > 400$ m/s with a 12-channel array would likely produce the best results. However, one would not know this until several different processing schemes with variable sub-array length had been investigated relative to a limited number of invasive CPT soundings.

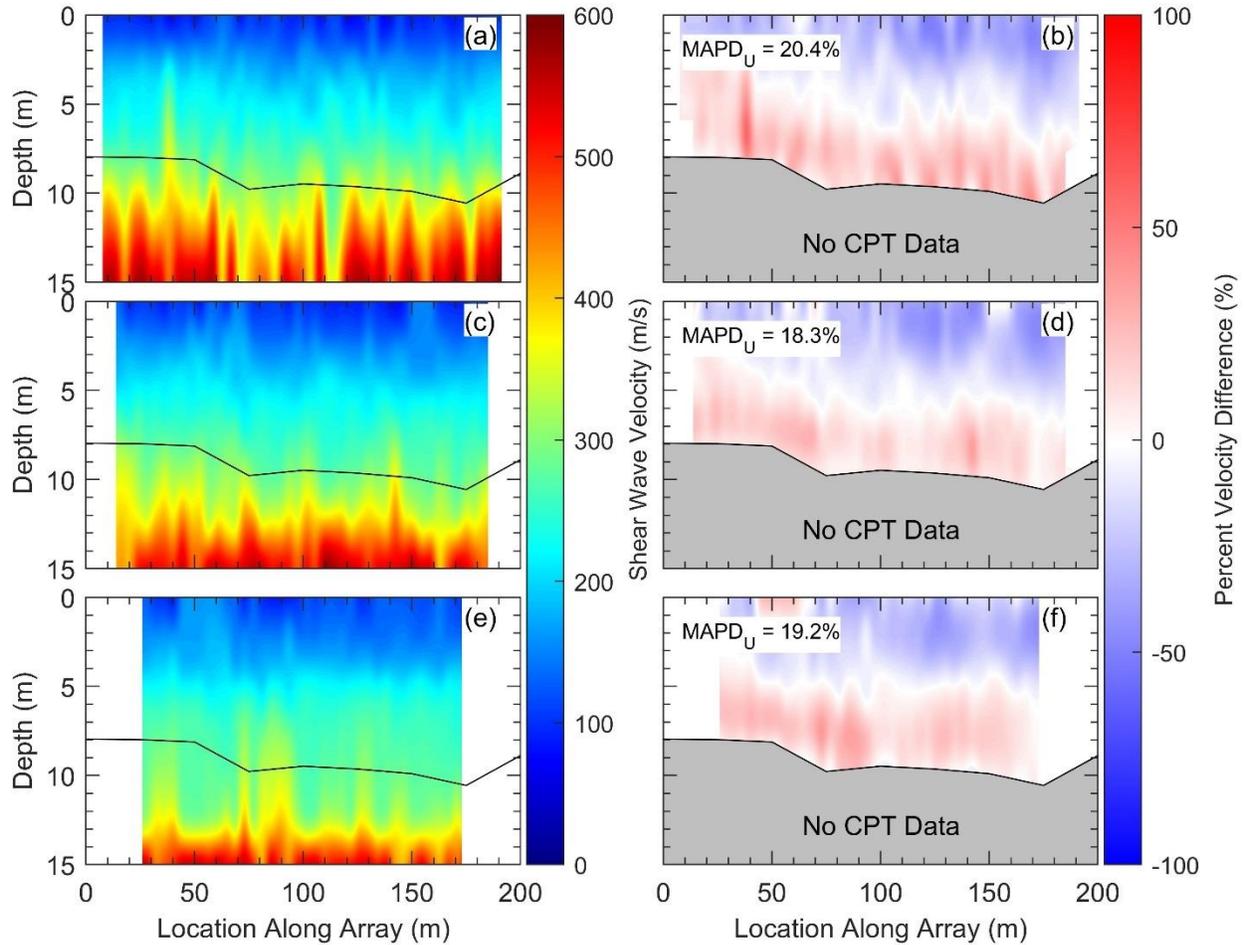

Figure 8: (a) Pseudo-2D $V_S$ cross-section from the 47, 12-channel MASW sub-arrays inverted using the U15 parameterization. (b) Comparison between the pseudo-2D $V_S$ cross-sections obtained from the 12-channel MASW sub-arrays and the CPT-to-$V_S$ correlation results. (c) Pseudo-2D $V_S$ cross-section from the 44, 24-channel MASW sub-arrays inverted using the U15 parameterization. (d) Comparison between the pseudo-2D $V_S$ cross-sections obtained from the 24-channel MASW sub-arrays and the CPT-to-$V_S$ correlation results. (e) Pseudo-2D $V_S$ cross-section from the 38, 48-channel MASW sub-arrays inverted using the U15 parameterization. (f) Comparison between the pseudo-2D $V_S$ cross-sections obtained from the 48-channel MASW sub-arrays and the CPT-to-$V_S$ correlation results. The mean absolute percent difference (MAPD) between the reference 2D image obtained from the CPT-to-$V_S$ correlation and the inversion results is displayed in the top left of each comparison plot. The depths of refusal for 9 CPT soundings along the array are shown on all plots with a solid black line.

In order to better quantify the variations in the 2D MASW $V_S$ cross-sections derived from various sub-array lengths, the CPT-to-$V_S$ correlation results presented in Figure 3 can be used as a common point of reference (at least for depths above CPT refusal). Again, we stress that using the CPT-to-$V_S$ correlation results as a common point of reference should not be interpreted as implying they represent the true values of $V_S$ within the subsurface. Rather, they simply provide a useful point of comparison for the various inversion results presented in this study. The percent differences between the $V_{S,CPT}$ values and the inverted $V_S$ values in the upper portions of the 2D MASW $V_S$ cross-sections above CPT refusal are shown in Figures 8b, 8d, and 8f for the 12- 24-, and 48-channel sub-arrays respectively. The cross-sections from all three sets of sub-arrays have negative percent difference values (i.e., lower $V_S$) at shallower depths and positive percent difference values (i.e., higher $V_S$) at depths closer to refusal when compared to the CPT results. To more quantitatively compare all three sets of results, the mean absolute percent difference (MAPD) in $V_S$ was calculated between each 2D MASW cross-section and the reference CPT cross-section. The MAPD is the mean value of the point-by-point absolute difference between each $V_S$ value and the reference $V_{S,CPT}$ value normalized by the reference $V_{S,CPT}$ value and expressed as a percentage. These MAPD values are shown in Figure 8 and are included in Table 1 along with the mean percent difference ($MPD_U$) expressed as a percentage. Note that these values all include a "$_U$" subscript (e.g., $MAPD_U$), to indicate they have been calculated only over the upper portion of the cross-section (i.e., above the CPT refusal line).

Both of these values provides different information about how each cross-section compares to the reference. The $MAPD_U$ values are very similar for all three sub-array lengths, ranging from 18.3% for the 24-channel results to 20.4% for the 12-channel results, indicating that all three 2D MASW $V_S$ cross-sections have roughly the same level of overall agreement with the CPT results, but not providing any information about the distribution of deviations in terms of magnitude or polarity. That information can be provided by the $MPD_U$ values, which are all relatively close to zero, indicating a balance between positive and negative differences. These results indicate that, quantitively, all three 2D MASW cross-sections agree with the CPT results to roughly the same degree and without any significant bias towards higher or lower $V_S$ relative to the CPT-correlated $V_S$ cross-section.

Table 1: Comparison of mean absolute percent difference (MAPD) and mean percent difference (MPD) statistics for the U15 cross-sections relative to the CPT-to-$V_S$ correlated reference cross-section. Note that the "U" subscript denotes the upper portion of the cross-section above the depth of CPT refusal, while the "L" subscript denotes the lower portion of the cross-section below CPT refusal.

| Upper: CPT reference | | |
|---|---|---|
| Channels | $MAPD_U$ (%) | $MPD_U$ (%) |
| 12 | 20.4 | -1.8 |
| 24 | 18.3 | -3.5 |
| 48 | 19.2 | -0.4 |
| Lower: 24-channel reference | | |
| Channels | $MAPD_L$ (%) | $MPD_L$ (%) |
| 12 | 14.9 | 7.7 |
| 24 | NA | NA |
| 48 | 12.7 | -9.4 |

Unfortunately, the limited depth of the CPT results makes it challenging to quantitatively compare the lower portions of the cross-sections at depths below the CPT refusal line (> ~ 8 - 10 m). To make these comparisons, a different reference condition needed to be chosen. Hence, the results from the 24-channel sub-arrays were used as the reference for the lower/deeper portions of the 2D $V_S$ images. Figure 9 shows the same U15 results and upper difference values as presented in Figure 8, with the addition of lower difference values for the portion of the cross-section below CPT refusal. The $MAPD_L$ and $MPD_L$ values for the 12- and 48-channel cross-sections are included in Table 1, with the $MAPD_L$ values also shown in Figure 9, to characterize the difference in the $V_S$ values in the lower portion of each cross-section. The $MAPD_L$ values are similar for both sub-array lengths, but $MPD_L$ values are very different, with 7.7% for the 12-channel results and -9.4% for the 48-channel results. This shows that, while the overall level of agreement and absolute magnitude of the differences are similar for both the 12- and 48-channel results, the 12-channel cross-section has more areas with higher $V_S$ while the 48-channel cross-section has more areas with lower $V_S$. These values are consistent with the trends discussed above: the shorter, 12-channel sub-arrays result in shallower impedance contrasts, leading to higher $V_S$ when compared to the 24-channels results. Conversely, the longer, 48-channel sub-arrays result in deeper impedance contrasts, leading to lower $V_S$.

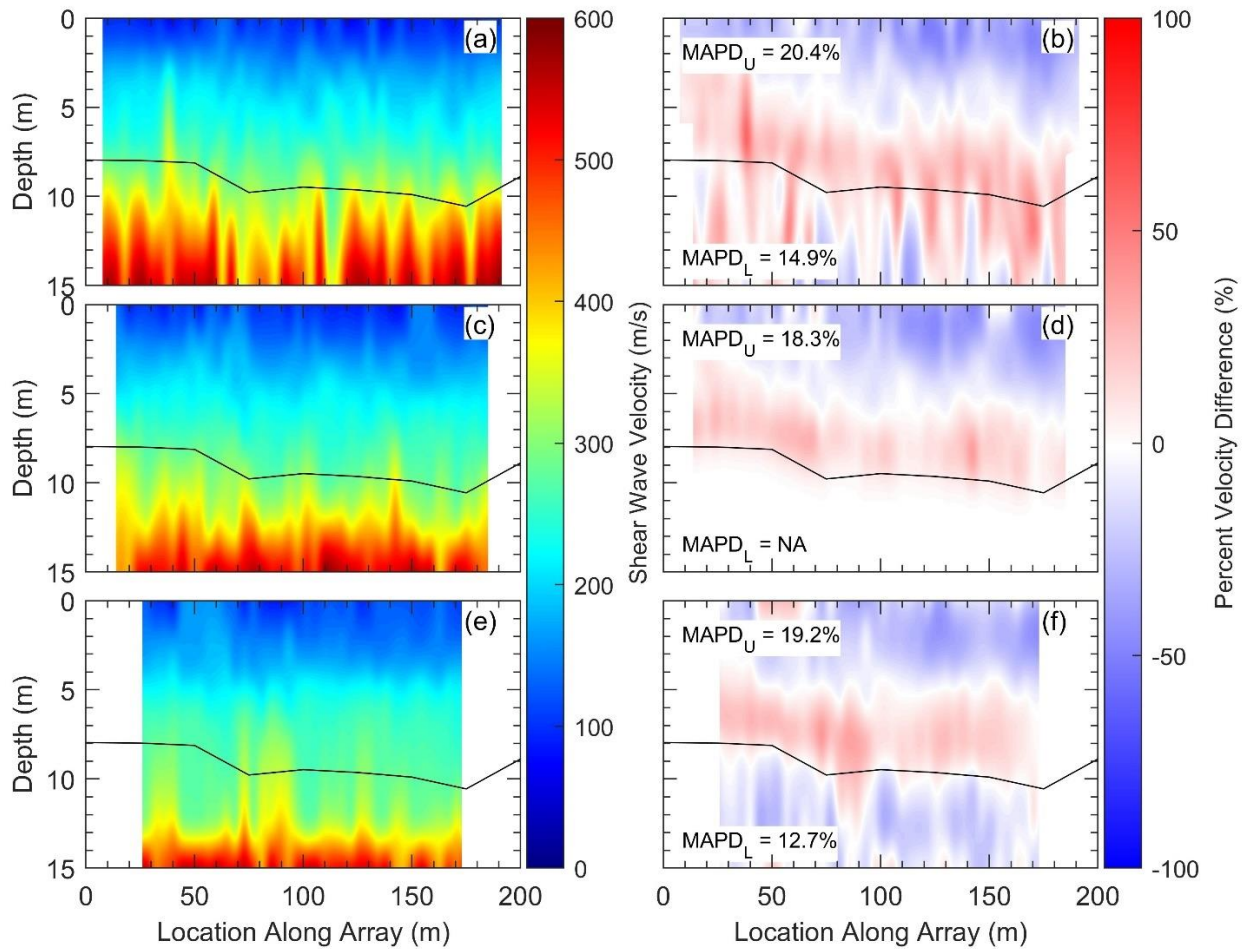

Figure 9: (a) Pseudo-2D $V_S$ cross-section from the 47 12-channel MASW sub-arrays inverted using the U15 parameterization. (b) Comparison between the pseudo-2D $V_S$ cross-sections obtained from the 12-channel MASW sub-arrays and (upper) the CPT-to-$V_S$ correlation results or (lower) the 24-channel MASW sub-arrays. (c) Pseudo-2D $V_S$ cross-section from the 44, 24-channel MASW sub-arrays inverted using the U15 parameterization. (d) Comparison between the pseudo-2D $V_S$ cross-sections obtained from the 24-channel MASW sub-arrays and (upper) the CPT-to-$V_S$ correlation results or (lower) the 24-channel MASW sub-arrays. (e) Pseudo-2D $V_S$ cross-section from the 38, 48-channel MASW sub-arrays inverted using the U15 parameterization. (f) Comparison between the pseudo-2D $V_S$ cross-sections obtained from the 48-channel MASW sub-arrays and (upper) the CPT-to-$V_S$ correlation results or (lower) the 24-channel MASW sub-arrays. The mean absolute percent difference values for the upper ($MAPD_U$) and lower ($MAPD_L$) portions are displayed in the top left and bottom left of each comparison plot, respectively. The depths of refusal for 9 CPT soundings along the array are shown on all plots with a solid black line.

One potential difference between these cross-sections is the uncertainty of the inversions used to produce them. While all surface wave inversion is uncertain due to the non-unique nature of dispersion data, the apparent amount of inversion-derived uncertainty is heavily influenced by the constraints applied to the inversion algorithm (Foti et al. 2015). These constrains include, the inversion target, which in this case is the experimental dispersion data that, as previously discussed, is influenced by the selected sub-array geometry, and the layering parameterization, which will be the focus of the present discussion. For the cross-sections shown in Figure 9, the U15 parameterization with its relatively-large number of layers may not be well constrained by the dispersion data due to the dispersion data's limited bandwidth for some of the 12- and 24-channel sub-arrays. As a result, it is not conclusive whether the variation in results in Figure 9, particularly at depth, is due to the sub-array length (our hypothesis) or the U15 parameterization. To investigate this issue further, the results from the inversions using LN parameterizations that have fewer layers (i.e., 3, 4, 5, and 6 layers) and which may be better constrained by the dispersion data from the 12- and 24-m sub-arrays are examined in the following discussion.

All three sets of MASW sub-arrays were inverted using the four LN parameterizations developed for this study (i.e., LN = 3, 4, 5, and 6). The resulting $V_S$ cross-sections from the 24-channel sub-arrays for each LN parameterization are shown in Figure 10. This figure demonstrates how the choice of layering parameterization can have a significant impact on the resulting $V_S$ cross-sections. The LN=3 cross-section has three distinct layers with more abrupt velocity contrasts between each layer. Conversely, the LN=6 cross-section has much less distinct layering, as the individual velocity contrasts between any two layers are more gradual. The LN=4 and LN=5 cross-sections display behaviors in between those of the LN=3 and LN=6 cross-sections. This presents a challenge to analysts who may wish to select a single $V_S$ cross-section for design purposes, or for further analysis such as determining the spatial variability of depth to rock or CPT refusal across the site. An analyst's first instinct may be to simply use the cross-section from the parameterization that has the lowest inversion misfits, but that would not be particularly helpful in this case, as all of the parametrizations had misfits that were very similar. The minimum misfits from each of the four sets of inversions ranged from 0.03 to 0.05, with the maximum misfits from each ranging from 0.30 to 0.32. The maximum and minimum misfits from the previously discussed U15 parameterization also fall within these ranges. These values indicate that none of the parameterizations appear to fit the experimental data any better or worse than any of the others. Therefore, an alternate approach must be utilized to determine which parameterization is most reasonable and/or which one is most useful for the intended purpose. As noted above in regards to the U15 inversions, some calibration to invasive field testing results may be needed in order to determine the best MASW processing and inversion scheme for a given application. For example, in regards to the results presented in Figure 10, if one were looking to determine the depth to CPT refusal across this site, using a criteria of $V_S > 400$ m/s with an LN = 3 layering parameterization would likely produce the best results. However, this would again require several different inversion layering parameterizations to have been investigated before a relationship could be established.

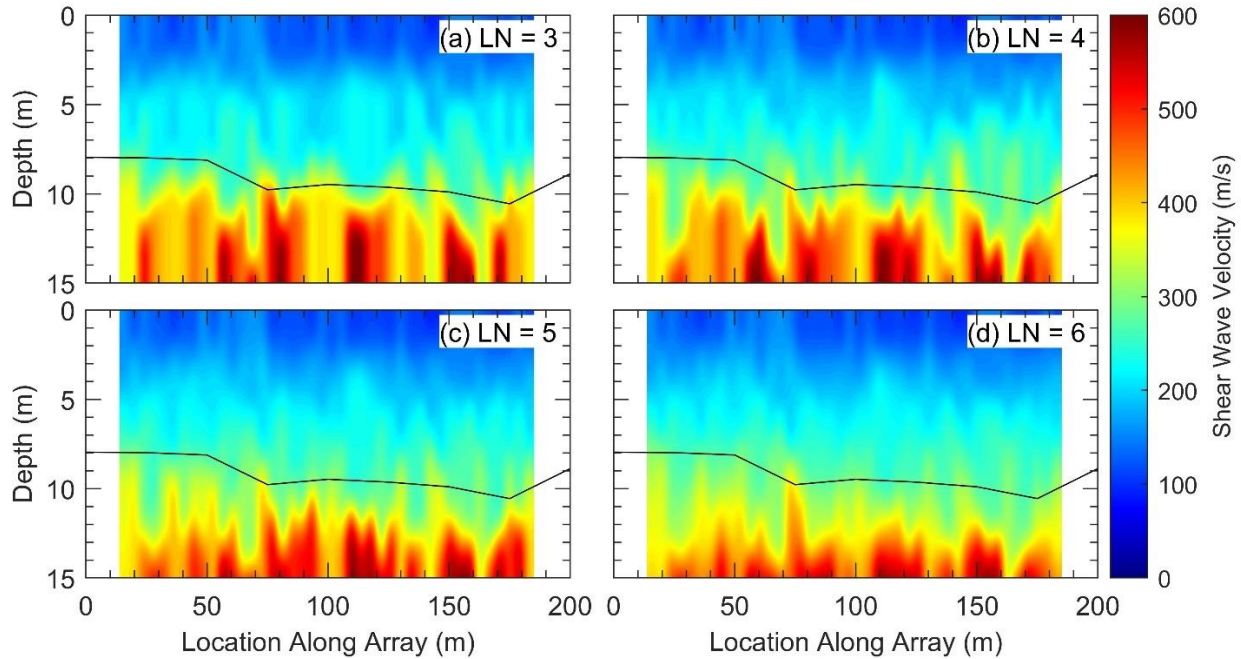

Figure 10: Pseudo-2D $V_S$ cross-sections from the 44, 24-channel MASW sub-arrays inverted using the (a) LN=3, (b) LN=4, (c) LN=5, and (d) LN=6 layering parameterizations. The depths of refusal for 9 CPT soundings along the array are shown on all plots with a solid black line.

If one is concerned about knowing the "true" layering and $V_S$ changes across a site, *a priori* information from boring logs, CPT data, or knowledge of local geology will be necessary to constrain the layering inversion (e.g., Teague et al. 2018). For example, at the site under consideration in this paper, the subsurface above 8 – 10 m is known to consist of 3 – 4 distinct layers based on CPT data (refer to Figure 2). However, the CPT data cannot inform about the layering at greater depths. Thus, if *a priori* information about the true subsurface layering is either not available, or does not extend to great enough depths, an approach like the DeltaVs method (Yust & Cox 2022) can be used to investigate the most likely layering beneath each sub-array. To accomplish this for the present datasets, the DeltaVs method was applied to all 1500 "best" profiles from each sub-array (i.e., 300 $V_S$ profiles for each of the five inversion parameterizations). The DeltaVs method uses the inverted $V_S$ profiles from all reasonable layering parameterizations to determine the likely number of significant layers in the subsurface below each sub-array. It does this by identifying clusters of layer boundaries within the large number of non-unique $V_S$ profiles resulting from surface wave inversions. The reader is referred to Yust & Cox (2022) for additional details on the application and limitations of the DeltaVs method. For the vast majority (84%) of the sub-arrays in the present study, the DeltaVs method indicated the presence of only three significant layers in the top 15 m, with four significant layers indicated for three to seven of the sub-arrays in each set and five significant layers indicated for only one or two sub-arrays in each set. The DeltaVs results for each sub-array were used to select the LN parameterization with the closest number of layers, which was, in most cases, LN=3, but also included some LN=4 and LN=5 results. Once the optimal LN parameterization was determined for each sub-array, the median $V_S$ profile for that parameterization was assigned to the mid-point of that sub-array, and spatial interpolation was used in the same manner described above to generate a single pseudo-2D $V_S$ cross-section for each set of sub-arrays. The pseudo-2D $V_S$ cross-sections obtained from layering optimization using the LN-DeltaVs procedure are shown in Figures 11a, 11c, and 11e for the 12-, 24-, and 48-channel sub-arrays, respectively. The corresponding comparisons relative to the reference conditions are shown in Figures 11b, 11d, and 11f, with the upper portion referenced to the $V_{S,CPT}$ values and the lower portion referenced to the 24-channel LN-DeltaVs results.

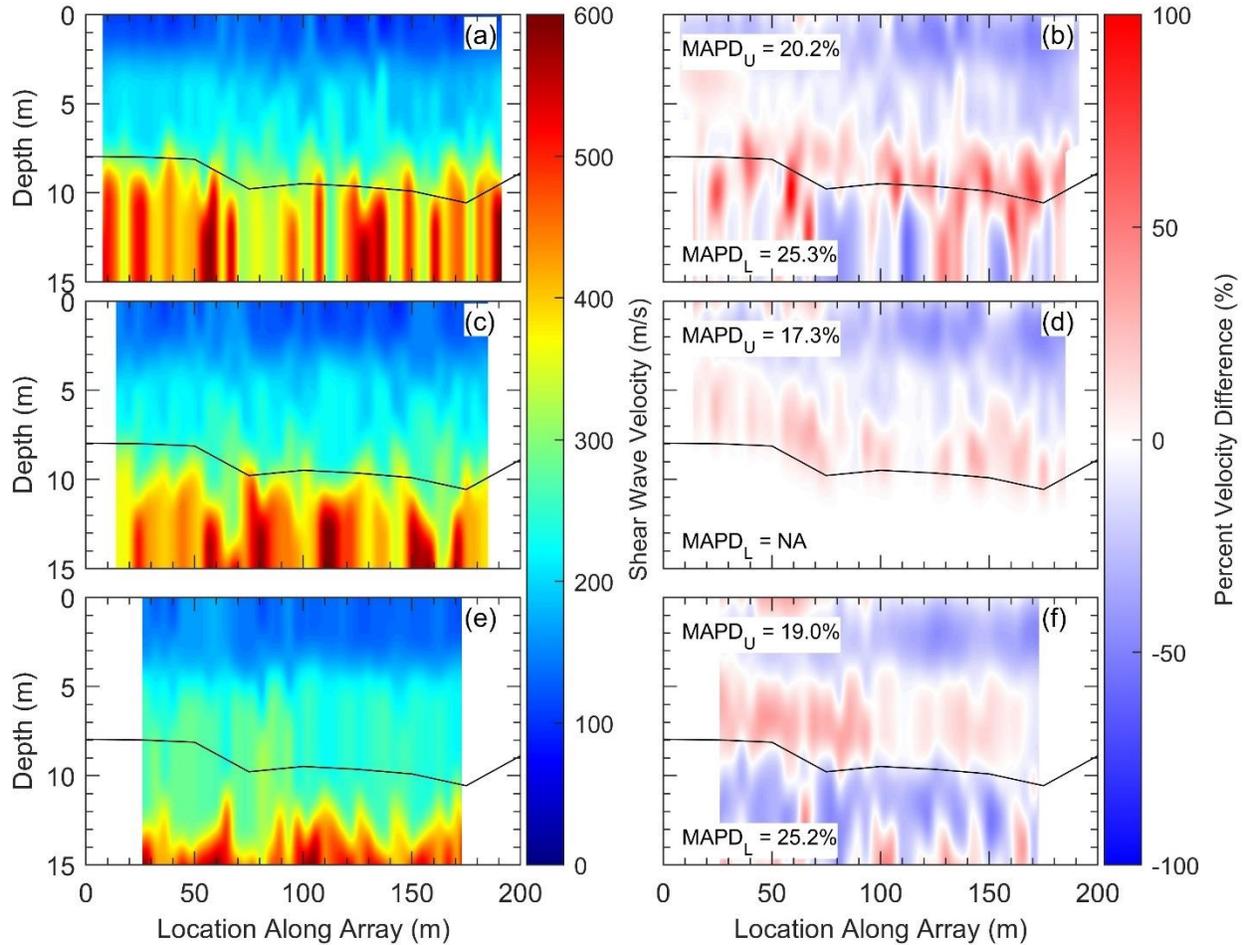

Figure 11: (a) Pseudo-2D $V_S$ cross-section from the 47 12-channel MASW sub-arrays inverted using the LN-DeltaVs parameterizations. (b) Comparison between the pseudo-2D $V_S$ cross-sections obtained from the 12-channel MASW sub-arrays and (upper) the CPT-to-$V_S$ correlation results or (lower) the 24-channel MASW sub-arrays. (c) Pseudo-2D $V_S$ cross-section from the 44, 24-channel MASW sub-arrays inverted using the LN-DeltaVs parameterizations. (d) Comparison between the pseudo-2D $V_S$ cross-sections obtained from the 24-channel MASW sub-arrays and (upper) the CPT-to-$V_S$ correlation results or (lower) the 24-channel MASW sub-arrays. (e) Pseudo-2D $V_S$ cross-section from the 38, 48-channel MASW sub-arrays inverted using the LN-DeltaVs parameterizations. (f) Comparison between the pseudo-2D $V_S$ cross-sections obtained from the 48-channel MASW sub-arrays and (upper) the CPT-to-$V_S$ correlation results or (lower) the 24-channel MASW sub-arrays. The mean absolute percent difference values for the upper (MAPD$_U$) and lower (MAPD$_L$) portions are displayed in the top left and bottom left of each comparison plot, respectively. The depths of refusal for 9 CPT soundings along the array are shown on all plots with a solid black line.

The LN-DeltaVs cross-sections have a roughly 3-layer structure, which is expected given that 84% of the median profiles selecting using the DeltaVs method were from the LN=3 inversion results. The locations of the layer boundaries within the top one-third of the cross-section are less erratic in the 48-channel cross-section than in the 12- or 24-channel cross-sections. This is, similar to the U15 results and consistent with the smearing effects discussed by Park (2005) and Mi et al. (2017). The behavior in the lower two-thirds of the cross-sections is also consistent with the U15 results, with longer sub-arrays resulting in higher velocities being pushed deeper. The average depths where velocities exceed 400 m/s are 11.49 m, 11.95 m, and 13.54 m for the 12-, 24-, and 48-channel cross-sections, respectively. These exceedance depths are very similar to those of the U15 results, with the only differences greater than 0.5 m being the 24-channel depth to 400 m/s, which is 0.81 m shallower than the results from the U15 inversions (refer to Figure 9).

In order to evaluate which cross-sections agree best with the CPT data and to better quantify the differences between the results from the various sub-arrays, the same MAPD and MPD statistics were calculated for the LN-DeltaVs results and are included in Figure 11 and Table 2. The LN-DeltaVs cross-sections agree with the CPT results slightly better than the U15 results, with reductions of $MAPD_U$ of between 0.2% and 1.0%. The $MPD_U$ values are all below zero, indicating a slight bias toward lower values of $V_S$ relative to the CPT correlations. This is more apparent in the 12- and 24-channel comparisons shown in Figures 11b and 11d, which appear to have more regions of lower $V_S$ in the upper portions of the cross-section above CPT refusal. Overall, while these values indicate that the LN-DeltaVs cross-sections have slightly better agreement with the CPT correlations and a small bias toward lower $V_S$ values, these results are very similar to those from the U15 cross-sections.

Table 2: Comparison of mean absolute percent difference (MAPD) and mean percent difference (MPD) statistics for the LN-DeltaVs cross-sections relative to the CPT-to-$V_S$ correlated reference cross-section. Note that the "U" subscript denotes the upper portion of the cross-section above the depth of CPT refusal, while the "L" subscript denotes the lower portion of the cross-section below CPT refusal.

| Upper: CPT reference | | |
|---|---|---|
| Channels | $MAPD_U$ (%) | $MPD_U$ (%) |
| 12 | 20.2 | -4.0 |
| 24 | 17.3 | -5.5 |
| 48 | 19.0 | -1.6 |
| Lower: 24-channel reference | | |
| Channels | $MAPD_L$ (%) | $MPD_L$ (%) |
| 12 | 25.3 | 7.7 |
| 24 | NA | NA |
| 48 | 25.2 | -13.1 |

The major differences in the statistics from the LN-DeltaVs and U15 cross-sections are found in the lower portions below CPT refusal. Here, the 12- and 48-channel cross-sections differ from the 24-channel cross-section to a much greater degree than in the U15 results, with $MAPD_L$ values both around 25%, almost double the 14.9% and 12.7% values for U15. The $MPD_L$ values, however, indicate that the average polarity of those differences is very different. The 12-channel cross-section has a slight bias toward higher velocities, with a positive but small value of $MPD_L$ = 7.7%, while the 48-channel cross-section is clearly biased toward lower velocities $MPD_L$ = -13.1%. While this is the same trend seen in the U15 results, it is more extreme here, with the 12- and 24-channel cross-sections being closer and the 48-channel one being much further from them. The consistency of these behaviors across both the LN-DeltaVs and U15 cross-sections suggests that the constraints imposed on the inversions by the parameterization, or lack thereof for U15, is not the cause of the variations. For the LN-DeltaVs results, the deepest layer boundary in each sub-array median profile occurs above $\lambda_{max}/2$ for 85% of the 12-channel sub-arrays and 93% of the 24-channel ones, as opposed to none of the 12-channel sub-arrays and only 34% of the 24-channel ones for the U15 results. Therefore, the movement of the deepest impedance contracts does not appear to be the result of parameterization uncertainty and/or placing layer boundaries below the optimal $d_{res} = \lambda_{max}/2$. Instead, this suggests that the variations are caused by changes in the target dispersion data and, by extension, the sub-array length.

While examining the LN-DeltaVs cross-sections has confirmed that the choice of sub-array length has a significant impact on 2D MASW results, it has not helped in determining which set of sub-arrays produce a cross-section that most reasonably characterizes the subsurface conditions at the site. The statistical comparisons between the CPT-to-$V_S$ correlations and the 2D MASW cross-sections are not particularly helpful here, as they show that all inversion results have very similar levels of agreement, despite clear visual differences in the $V_S$ cross-sections. The depths to CPT refusal do visually agree best with the 12- and 24-channel cross-sections, but that is not enough to conclusively say which sub-array length produces the most reasonable results, as the most significant variations between cross-sections occur below the depth of CPT refusal. To address this problem, two boreholes were drilled at 12.5 m and 137.5 m along the alignment of the DAS array. The borehole at 12.5 m (B1) was drilled to 24.4 m (80 ft) and the borehole at 137.5 m was drilled to 15.2 m (50 ft), so they both fully cover the depth range between CPT refusal and the bottom of the cross-sections. Sampling was performed in both boreholes every 1.5 m (5 ft) and the collected material was classified according to the Unified Soil Classification System (ASTM 2017). The lithographic logs for both boreholes are shown in Figures 12a and 12b, along with the LN-DeltaVs cross-section from the 12- and 48-channel MASW sub-arrays. Additionally, casing was installed in the first borehole, B1, and downhole testing was performed by the NHERI@UTexas team down to a depth of 24 m with a receiver interval of 1 m. The NHERI@UTexas team identified four velocity layers from the downhole testing, which are also shown in Figures 12a and 12b to the left of the B1 lithographic log using the same color scale as the $V_S$ values of the cross sections.

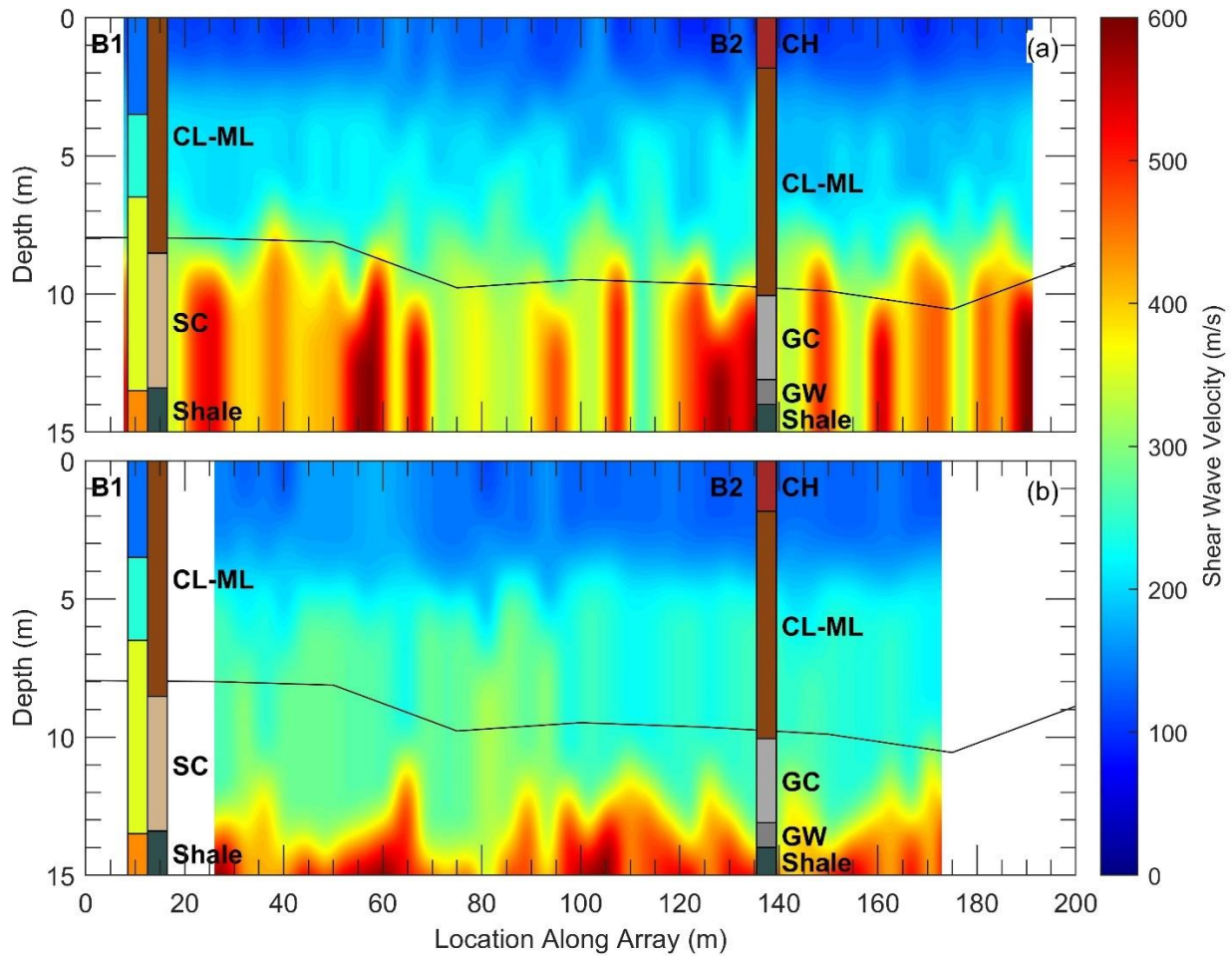

Figure 12: Pseudo-2D $V_S$ cross-section from the (a) 47, 12-channel and (b) 38, 48-channel sub-arrays inverted using the LN-DeltaVs parameterizations and the lithographic logs for boreholes B1 and B2 shown at their respective locations along the array. The four velocity layers from downhole testing in borehole B1 are shown next to the lithographic log for that borehole. The depths of refusal for 9 CPT soundings along the array are shown with a solid black line.

The logs from both boreholes are similar, with both showing a mix of granular and cohesive materials down to about the depth of CPT refusal, where they transition to denser granular materials before encountering a shale layer near a depth of 13.5 m. With the exception of the thin fat clay (CH) layer at the very top of log B2, both boreholes show the same sandy silty clay (CL-ML) layer for the top 8.5-10 m of depth. This layer is noted on the boring log reports as alternatively being classified as silty clayey sand (SC-SM), which is consistent with the silt and sand mixes characterized by the CPT soil behavior types for most of these depths. The bottom depth of the CL-ML layer (8.5 m for B1 and 10 m for B2) also agrees well with the depths to CPT refusal. Below this point, B1 transitions to dense clayey sand (SC), while B2 encounters dense clayey gravel with sand (GC) and well-graded gravel with sand (GW) layers. Finally, after the dense granular layers, both boreholes encounter a shale layer at depths of 13.4 m for B1 and 14 m for B2. The downhole results from B1 show a 3.5-m thick layer at the surface with $V_S$ = 140 m/s followed by a second layer with $V_S$ = 248 m/s down to 6.5 m and a third layer with $V_S$ = 359 m/s down to a depth of 13.5 m. A final layer with $V_S$ = 445 m/s was identified starting at 13.5 m and continuing to the bottom of the borehole at 23 m. While the layer boundaries at the bottoms of the first and second layers do not line up cleanly with the material boundaries identified in the lithographic logs, the boundary at the bottom of the third layer agrees very well with the top of the shale layer identified in both logs.

The depths of the top of the dense granular layers agree well with both the depths of CPT refusal and the shallower impedance contrasts shown in the 12- and 24-channel cross-sections. On its own, this would support the conclusion that the two sets of shorter sub-arrays produce $V_S$ cross-sections that are better representations of the subsurface conditions than that from the 48-channel sub-arrays. However, the shale layer found in both boreholes must also be considered. The transition to this material also corresponds to a significant $V_S$ impedance contrast in the downhole results, which agrees well with the deeper, high $V_S$ boundary in the 48-channel cross-section that occurs at shallower depths in the other two. It is not necessarily surprising that this layer does not show up at all in the 12-channel cross-section, as those sub-arrays did not resolve dispersion data with wavelengths long enough to accurately characterize that deep. For the 24-channel sub-arrays, 66% of them resolved wavelengths long enough to constrain this layer, which may explain why it appears like some of the profiles in that cross-section place the impedance contrast lower, while others have it up near CPT refusal. The 48-channel sub-arrays all have dispersion wavelengths that were great enough to constrain a layer at approximately 14 m, likely contributing to its prominent appearance in the 48-channel cross-section.

These results emphasize the importance of using invasive data to calibrate the processing and inversion procedures based on the desired information or application. As noted above, if one desired to identify the depth to CPT refusal and the dense granular layers seen in the boreholes, the best option would be to identify the depths at which $V_S$ exceeds 400 m/s in the 12-channel sub-array results shown in Figure 12a. Alternatively, if one's goal was to identify the depth of the shale layer, applying the same criteria ($V_S$ > 400 m/s) to the 48-channel sub-array results shown in Figure 12b would yield the best results. Conversely, if locating the top of shale were the desired testing outcome and 12-channel sub-arrays were selected prior to acquisition using the roll-along method, the results would have been misinterpreted significantly. Additionally, the 48-channel results agree best with the $V_S$ measurements from the downhole test at B1, suggesting that the 48-channel cross-section is the one most likely to resemble the true subsurface conditions. The shorter sub-arrays are better able to characterize the shallower portions of the subsurface, identifying layers that would not be found with longer-sub arrays, while, somewhat unsurprisingly, the longer sub-arrays improve characterization at depth. These factors are all in addition to the effects sub-array length has on horizontal resolution (i.e., lateral smearing) which were discussed above and may contribute to longer sub-arrays' inability to discern some shallower layer boundaries.

These results show that it is important to keep in mind the goals of a specific project when deciding what sub-array geometry to use for 2D MASW. They also suggest that to address multiple goals or fully characterize the subsurface conditions at the site, it may be necessary to use some combination of different sub-array geometries calibrated to various *a priori* information. While this is possible with traditional roll-along geophone surveys, it is inconvenient to have to alter the physical setup substantially for different sites or perform multiple surveys at the same site. Additionally, if there is no *a priori* information about the site, multiple sub-array geometries need to be considered for robust characterization, which is highly impractical with traditional equipment. In both cases, the ability of DAS to allow the analyst to select the sub-array geometry, or geometries, during the processing stage rather than needing to predetermine the survey geometry before data acquisition makes DAS a much more efficient method for performing pseudo-2D MASW imaging. Overall, as much data as possible should be collected in the field to allow for flexible sub-array processing after the fact, whether through the used of larger geophone arrays or DAS. By collecting the data required for flexible sub-array processing, analysts are able to properly consider sub-array geometry as a potential source of epistemic uncertainty and, when possible, calibrate the analysis based on project goals and *a priori* information.

## Conclusions

This study examined the effects of 2D MASW sub-array length using DAS data collected at a well-characterized site. Specifically, three sets of sub-arrays of varying lengths were used to develop pseudo-2D $V_S$ cross-sections along the same 200-m long DAS line: (a) 12-channel sub-arrays approximately 11-m long, (b) 24-channel sub-arrays approximately 23-m long, and (c) 48-channel sub-arrays approximately 47-m long. An equivalent sub-array spatial interval equal to 4-channels (approximately 4 m) was used for all sub-arrays, totaling 129 individual MASW analyses. Rigorous dispersion processing and inversion techniques (e.g., dispersion uncertainty from multiple shot locations and investigating multiple inversion layering parameterizations to quantify epistemic uncertainty) not used in previous, documented applications of 2D MASW were used to characterize the subsurface as comprehensively as possible while accounting for multiple sources of uncertainty. The $V_S$ cross-sections resulting from each set of sub-arrays were visibly different from one another, illustrating how sensitive the results of 2D MASW can be to the choice of sub-array length. However, different cross-sections were shown to better correlate with different key features of the subsurface, as determined from comparisons with invasive data collected along the array alignment, including correlated $V_S$ values from CPT soundings, borehole lithography logs, and downhole seismic testing results.

For this site, inverting dispersion data from longer sub-arrays resulted in significantly deeper impedance contrasts closer to the bottom of the characterizable depth, which agreed well with the location of a shale layer found in the borehole logs, while shorter sub-arrays resulted in shallower impedance contrasts that compared favorably with the depths of CPT refusal and locations of dense granular layers found in the boreholes. The longer sub-array results also agreed best overall with the downhole velocity measurements taken in one of the boreholes. These results further support the conclusion that *a priori* information, such as invasive testing results, should be used to calibrate the analysis procedures to specifically address project-specific goals, when possible. Further, in the absence of *a priori* information that conclusively constrains subsurface layering, analysts may need to consider multiple 2D MASW sub-array configurations to comprehensively characterize the subsurface conditions and properly evaluate the uncertainty of the 2D MASW results, which has not been considered by previous studies on 2D MASW in the field.

This study also demonstrates the abilities of using DAS data to perform pseudo-2D MASW imaging and highlights the advantages of DAS over traditional equipment when it comes to flexibility in changing processing parameters like sub-array length and channel separation to address project-specific goals. The ability of DAS to collect high spatial resolution records simultaneously along significant lengths of fiber optic cable for each shot allows for data acquisition at rates much faster and more efficiently than traditional roll-along geophone surveys (at least once the cable has been installed), while allowing the sub-array length, array interval, channel separation, and processed shot locations to be selected or changed after acquisition during the processing stage. As this and previous studies have found that lateral resolution, maximum characterization depth, and anomaly detection are all heavily influenced by sub-array length, DAS allows for data to be reprocessed depending on the areas and features of interest for the individual project. This is not possible with traditional 2D MASW equipment, as the sub-array length, sub-array interval, and shot location must all be pre-determined for the acquisition stage and cannot easily be changed during processing. Overall, DAS is a promising new technology for recording surface-wave data that is very well suited to 2D MASW testing. While some existing limitations of the method persist when utilizing

DAS, as they are tied to the inherently 1D nature of MASW analysis, DAS does allow for 2D MASW to be performed in a much more rigorous, efficient, and flexible manner than traditional surveys.

## Acknowledgments

This work was supported by the U.S. National Science Foundation (NSF) Graduate Research Fellowship under Grant No. DGE-2137420 and by NSF grants CMMI-2037900, CMMI-1520808, and CMMI-1931162. However, any opinions, findings, conclusions, or recommendations expressed in this material are those of the authors and do not necessarily reflect the views of NSF. Special thanks to Dr. Kevin Anderson at Austin Water – Center for Environmental Research for the access to the Hornsby Bend Biosolids Management Plant test site. Special thanks to Dr. Kenichi Soga for the contribution of the NanZee cable used in this study.